\documentclass[aps,prx, amsmath,amssymb,reprint,groupedaddress]{revtex4}

\usepackage{graphicx}% Include figure files
\usepackage{dcolumn}% Align table columns on decimal point
\usepackage{bm}% bold math

\usepackage{hyperref}

\newcommand{\cmmnt}[1]{\ignorespaces}

\begin{document}

\title{Phase-matching mechanism for high-harmonic generation in the overdriven regime driven by few-cycle laser pulses}

\author{J. Sch\"otz$^{1,2}$}
 \email{johannes.schoetz@mpq.mpg.de}
\author{B. F\"org$^{1,2}$}
\author{W. Schweinberger$^{1,3}$}
\author{I. Liontos$^{3}$}
\author{H.A. Masood$^{3}$}
\author{A. M. Kamal$^{3}$}
\author{C. Jakubeit$^{2}$}
\author{N. G. Kling$^{1}$}
\author{T. Paasch-Colberg$^{1,3}$}
\author{M. H\"ogner$^{1,2}$}
\author{I. Pupeza$^{1,2}$}
\author{M. Alharbi$^{3}$}
\author{M.F. Kling$^{1,2}$}
 \email{matthias.kling@lmu.de}
\author{A.M. Azzeer$^{3}$}
 \email{azzeer@ksu.edu.sa}

\affiliation{$^1$ Physics Department, Ludwig-Maximilians-Universit\"at Munich, D-85748 Garching, Germany}
\affiliation{$^2$ Max Planck Institute of Quantum Optics, D-85748 Garching, Germany}
\affiliation{$^3$ Attosecond Science Laboratory, Physics and Astronomy Department, King Saud University, Riyadh, Saudi Arabia}

\date{\today}

\begin{abstract}
Isolated attosecond pulses (IAPs) produced through laser-driven high-harmonic generation (HHG) hold promise for unprecedented insight into biological processes via attosecond x-ray diffraction with tabletop sources. However, efficient scaling of HHG towards x-ray energies has been hampered by ionization-induced plasma generation impeding the coherent buildup of high-harmonic radiation. Recently, it has been shown that these limitations can be overcome in the so-called 'overdriven regime' where ionization loss and plasma dispersion strongly modify the driving laser pulse over small distances, albeit without demonstrating IAPs. Here, we report on experiments comparing the generation of IAPs in argon and neon at 80\,eV via attosecond streaking measurements. Contrasting our experimental results with numerical simulations, we conclude that IAPs in argon are generated through ionization-induced transient phase-matching gating effective over distances on the order of 100\,$\mu$m. We show that the decay of the intensity and blue-shift due to plasma defocussing are crucial for allowing phase-matching close to the XUV cutoff at high plasma densities. We perform simulations for different gases and wavelengths and show that the mechanism is important for the phase-matching of long-wavelength, tightly-focused laser beams in high-pressure gas targets, which are currently being employed for scaling isolated attosecond pulse generation to x-ray photon energies.
\end{abstract}

%Use showkeys class option if keyword
%display desired
\maketitle

\flushbottom
\section{Introduction}

The interaction of intense laser pulses with atoms, where the laser electric field strength is comparable to the Coulomb field within an atom, leads to the generation of very high order harmonics of the fundamental photon energy\cite{McPherson1987, Huillier1988, Huillier1989}. High-harmonic generation can be understood in an intuitive semi-classical model\cite{Corkum93threestep}, which describes the process in three steps: An electron tunnels out of the atom through the potential barrier lowered by the strong electric field of the laser pulse. Subsequently, it propagates under the influence of the electric field. Finally, the electron recollides and recombines with the atom, which generates a photon whose energy is given by the sum of the kinetic energy of the electron and the ionization potential of the atom. The photon energy typically lies in the extreme ultraviolet (XUV, 10\,eV$\sim$100\,eV) or x-ray region (100\,eV$\sim$1\,keV).

Fueled by advances in the sub-cycle waveform control of laser pulses \cite{steinmeyer1999frontiers,brabec2000intense}, the exploitation of the intrinsic synchronization between the HHG radiation and the driving laser field led to the development of the field of attosecond physics\cite{Agostini2001firstRABBIT,Krausz2001first_isolated,Krausz2009RevModPhys,Keller2012RevAttophysics,Calegari2016Attoreview,Tzallas2019AttosecondPulseMetrology}. The generation of isolated attosecond pulses extended the powerful techniques of pump-probe spectroscopy to the electronic time scale and paved the way to a plethora of applications. IAPs were first demonstrated using attosecond streaking\cite{Itatani2002attosecondStreak, kitzler2002quantum, Goulielmakis_2004_first_streaking, kienberger2004_streaking}, a laser-assisted photoelectron spectroscopy technique which allows the reconstruction of both the IAP and the probe laser field. It is based on the time-delay dependent change of the final energy of the photoelectrons due to the propagation in the strong probe laser field after the emission process.

Generally, attosecond bursts are generated during each half-cycle of a laser pulse. In order to extract single isolated attosecond pulses from HHG radiation, several gating schemes have been proposed and demonstrated experimentally\cite{Calegari2016Attoreview}. Amplitude gating relies on spectral selection of the cutoff, which is generated only during the central most intense half-cycle of the few-cycle generating field\cite{Krausz2001first_isolated}. Furthermore, the efficiency of the recollision process can be controlled by the ellipticity of the laser pulse and effectively confined to a single field oscillation, forming the basis of polarization gating\cite{Nisoli2006SciencePolarization} and double optical gating\cite{Chang09DOG}, which additionally introduces a weak second harmonic field. Finally, ionization of the gaseous medium provides two more ways to generate IAPs. First, by using a strong enough field, the ground state of the atom can be depleted by a single half-cycle in the leading edge of a laser field, inhibiting XUV emission during later half-cycles, which results in a single IAP\cite{Leone2007generating, ferrari2010ionization}. Secondly, the ionization of atoms leads to an additional time-dependent dispersive plasma term for the driving laser pulse, such that phase-matching between the driving pulse and generated HHG can be confined to a single half-cycle\cite{Agostini1999optimizing, Balcou2003global,KapteynMurnane2006generation, Leone2009isolated, KapteynMurnane2009characterizing, KapteynMurnane2014generation}(transient phase-matching). The latter has recently sparked a lot of interest in HHG and especially single IAP generation with long-wavelength driving laser pulses that can reach up to 1\,keV photon energies\cite{Popmintchev2012xrayreview, Biegert2016soft_xray_continua}.

The balance of plasma-induced dispersion and atomic dispersion leading to perfect phase-matching is reached at the critical ionization fraction $\eta_{\mathrm{cr}}$ which, for slowly varying driving laser pulses, depends solely on the wavelength\cite{Popmintchev2009PhaseMatchingCutoffs}. For a given laser pulse shape and under the assumption that the driving laser changes only slowly within the target, this determines the maximum intensity for which phase-matching at the peak of the pulse is possible. This limits the maximum XUV energy and has been termed phase-matching cutoff. 

To increase the cutoff, longer wavelength driving lasers are required and in order to compensate for the smaller single atom generation efficiency, high-pressure gas targets are employed\cite{Biegert2016soft_xray_continua,Kaertner2011plasmadefocusing,Popmintchev2009PhaseMatchingCutoffs}. Higher gas pressures will ,however, lead to a higher plasma density and consequently reshaping of the driving laser pulse which affects HHG phase-matching for high intensity lasers and has been the subject of a number of theoretical\cite{Tempea2000NonadiabaticSelfPhasematching, Yakovlev2007enhancedPM, Liu2010theoHHGnonadiabaticPropagationIsolated, Kaertner2011plasmadefocusing} and experimental studies\cite{Vozzi2011PMhighenergyMIR,Chen2017plasmadefocusingExp,Tisch2018high}. Rapid increase of plasma density between consecutive half-cycles was predicted to strongly modify the driving laser through a blueshift, counteracting via the HHG dipole phase the phase-mismatch due to free electrons\cite{Tempea2000NonadiabaticSelfPhasematching}. This leads locally and transiently to nearly perfect phase-matching and a buildup of the XUV radiation over propagation lengths of 10-100\,$\mu$m. This mechanism was cited by several older experimental works that demonstrated XUV generation in tight focusing conditions significantly above the phase-matching cutoff up to the keV range\cite{Krausz1998coherentXray0p5keV, Spielmann2004xrayTitanium, Krausz2005sourceofcoherentxrays, Spielmann2006XAS_keVHHG}. It was later shown theoretically that this regime can also be achieved through the plasma induced intensity decay of the driving laser pulse in less extreme scenarios\cite{Yakovlev2007enhancedPM} and called self phase-matching (SPM). Transient phase-matching gating of IAPs based on SPM has been observed numerically\cite{Liu2010theoHHGnonadiabaticPropagationIsolated}. HHG with mid-infrared driving lasers in high-pressure gas targets with strong plasma-induced defocussing has been studied numerically and intensity clamping of the driving laser has been observed\cite{Kaertner2011plasmadefocusing}. The analysis of the phase-mismatch, however, only included the intensity decay in the dipole term, but not the blueshift.

A similar phase-matching expression has been used to explain experiments with a few-cycle 1.5\,$\mu$m driving laser pulse that observed enhanced HHG when placing the target in front of the focus\cite{Vozzi2011PMhighenergyMIR}. In their simulations they found that plasma defocussing plays an important role and that phase-matching can act as a transient gating mechanism, limiting the XUV generation to a few half-cycles.
Another experiment using strongly focused multicycle driving lasers centered at 800\,nm showed that the XUV cutoff can be significantly above the phase-matching cutoff and SPM has been used to explain the result qualitatively\cite{Chen2017plasmadefocusingExp}. One recent study that used sub-two cycle pulses at 1800\,nm could measure the carrier-envelope phase (CEP) dependence of the generated XUV and moreover could observe the plasma density experimentally\cite{Tisch2018high}. The simulations showed XUV buildup over few 100\,$\mu$m and an intensity clamping and plasma-induced pulse reshaping of the driving laser, and the term 'overdriven limit' has been used to describe this regime. All the experiments above, however, only measured the XUV spectra, only qualitatively analyzed the phase-matching, and did not directly temporally characterize the IAPs and driving laser pulses. 

Here, we study the generation of isolated attosecond pulses in neon and argon, respectively, at 80\,eV photon energies driven by few-cycle pulses at 750\,nm central wavelength and intensities around $5\cdot10^{14}$ W/cm$^2$, both experimentally and numerically. This is above the classical phase-matching cutoff for argon. The isolated attosecond pulses are generated via HHG in a gas target and subsequently fully characterized by attosecond streaking together with the few-cycle driving laser pulses. While the results in neon are in agreement with the employed amplitude gating scheme, we find that in argon the generation of isolated attosecond pulses cannot be understood from the measured laser pulses within the single atom picture, which hints at phase-matching effects. Numerical simulations are presented which shed light on the generation mechanism and are able to reproduce our experimental observations to a large extent. The simulations suggest that for argon our experiments are in the overdriven regime and that the driving laser pulses are strongly modified by ionization in the gas target. HHG is confined to the first few 100\,$\mu m$ and plasma-induced transient phase-matching is responsible for the gating of the isolated attosecond pulses. By comparison with numerical calculations, we show that only an extended analytic phase-matching expression that includes the strong plasma-induced driving pulse reshaping, including intensity decay and the blueshifting, is able to capture the phase-mismatch. Moreover, we present simulations for different wavelengths and gas types under strong focusing conditions and show that pulse reshaping contributes significantly to phase matching when the flux is maximized. Our findings are of relevance to the current efforts to push HHG into the water window and beyond.

\begin{figure*}[htb!]
	\centering\includegraphics[width=6.5in]{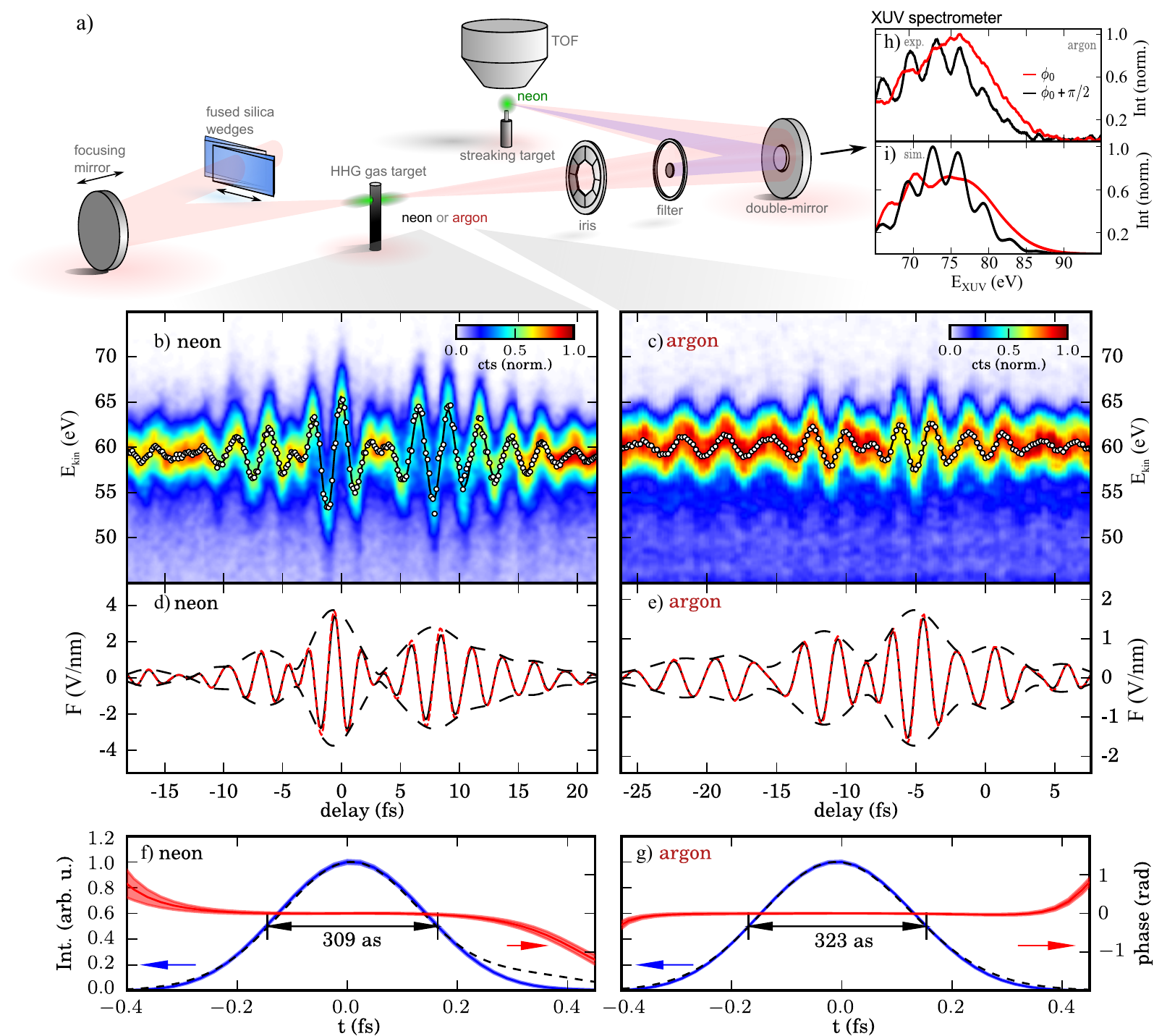}
	\caption{HHG generation at 80\,eV with sub-two-cycle pulses at 750\,nm in neon and argon: a) Scheme of the experimental setup. b) and c) Experimental attosecond streaking traces for isolated attosecond pulse generation through HHG in neon and in argon, respectively. The retrieved raw (dots) and filtered (solid line) streaking curve from the Gaussian fits is shown. d) and e) Electric field reconstructed from the streaking curves from the Gaussian fit (red dashed line) and the ptychographic reconstruction algorithm (black solid line). f) and g) reconstructed XUV pulse intensity (blue) and phase (red). The black dashed line shows the pulses from the simulation. The shaded areas show the standard deviation (see text for details). h) and i) Experimental and simulated XUV spectra in argon when the double mirror is replaced with a grating spectrometer for two different values of the CEP (see methods for details).
		\label{Fig:Spectrogram}}
\end{figure*}

\section{Results and discussion}
\subsection{Experimental pulse reconstruction}
\label{sec::discussion}

A scheme of the experimental setup is shown Fig. \ref{Fig:Spectrogram} a). In short, few-cycle laser pulses covering a wavelength range from 450\,nm to 1000\,nm and a central wavelength of 750\,nm are passing through a pair of fused silica wedges for dispersion tuning and are subsequently focused through the gas target for HHG. The gas target consists of a hollow metal tube with predrilled entrance and exit holes and is kept at a constant gas pressure of few 100\,mbar of either neon or argon, by controlling the gas flow. After passing through an iris the beam falls onto a concentric filter, which separates the beam into the XUV beam in the center and the laser beam in the outer region. A two-component mirror focuses the beams into a neon gas jet produced by the streaking target, and a time-of-flight-spectrometer (TOF) is used to record the kinetic energy of the generated photoelectrons. The inner part of the double mirror has a special multilayer coating to reflect XUV in a range of 6.4\,eV with a flat spectral phase around a central energy of 80\,eV and can be translated by a piezo stage in order to generate a variable time delay between the XUV and the laser pulse. Moreover, the double mirror can be removed such that the beam falls onto an XUV spectrometer (for further information on the laser system and experimental setup see methods and SI). We use the gas pressure, HHG focusing mirror position, and fused silica wedge insertion to optimize IAP generation for both cases (see methods and SI for details). We end up with similar gas pressure, an additional wedge insertion of 260\,$\mu$m and a focal mirror shift of roughly 5.5\,cm towards the gas target. 

The attosecond streaking spectrograms are shown in Fig. \ref{Fig:Spectrogram} for XUV generated in neon (b) and argon (c). In the delay dependent photoelectron spectrum clear oscillations of the mean electron energy can be observed. Both spectrograms are normalized, note, however, that the photoelectron count rate from argon is roughly an order of magnitude lower than from neon. From the streaking spectrogram, precise information about the waveform of the IR pulses in the streaking focus can be obtained as the shift of the mean photoelectron energy for each time delay essentially samples the laser pulse vector potential\cite{Itatani2002attosecondStreak}. For a coarse reconstruction, a streaking curve is extracted from the central energy of the fit of a Gaussian function to the photoelectron spectrum for each delay step (white dots). This curve is subsequently smoothed by filtering out frequency components below 450\,nm and above 1050\,nm (black line). From this the electric field of the laser pulse in the focus can be calculated\cite{schotz2017reconstruction} as is shown in Fig. 2 (d) and (e) for neon and argon, respectively (red dashed line). 

In order to reconstruct the attosecond XUV-pulses we employed the ptychographic reconstruction algorithm described in Ref. \cite{Lucchini2015ptychographic}.  As can be seen in Fig. \ref{Fig:Spectrogram} (f) and (g) the reconstruction yields XUV pulses (blue line) with 309$\pm $9\,as and 323$\pm$4\,as duration (FWHM), respectively, with almost negligible chirp and almost no satellite pulses (contrast better than 50:1) as is expected from the shape of the spectrograms (see the SI for further information on the reconstruction). The algorithm also returns the streaking laser pulse as shown in Fig. \ref{Fig:Spectrogram} (d) and (e) in black. Both streaking pulse shapes agree very well with the ones obtained through the Gaussian fits.

Note that the change in laser pulse shape is in a large part due to the spectral phase change caused by the different amount of fused silica wedge insertion (+260\,$\mu$m for argon). An additional small change in the waveform might hint at nonlinear pulse propagation within the argon gas target, but might also have a few other reasons and is therefore discussed in the SI. However, a subtle but very important difference of the pulse shapes can clearly be observed: For neon the electric field peaks with the envelope, while for argon the waveform yields a zero crossing at the peak of the envelope. Usually only the former is considered to allow for isolated attosecond pulse generation through amplitude gating. Indeed, this is a strong indication that a transient gating mechanism is at play in the isolated attosecond XUV pulse generation in argon.

\subsection{Simulation of pulse propagation and XUV generation}

In order to gain insight into the pulse propagation dynamics in the HHG target, we performed numerical simulations with a model described in more detail in Ref. \cite{hogner2017generation}. In short, for both the driving field and the XUV field linear refraction and absorption is considered in the paraxial approximation and rotational symmetry is employed. Moreover, for the driving laser field the Kerr effect and plasma generation, leading to ionization loss, plasma defocusing and blueshift, are taken into account. A static ionization rate that accounts for depletion is used\cite{Tonglin2005JourPhysB}. For the XUV emission, the dipole response of individual atoms is modeled using the strong-field approximation (SFA) including ground state depletion. After the HHG target, both beams are propagated through the beamline in the paraxial approximation (see SI for details).

The input pulse for neon is obtained from the measured streaking trace, subtracting the effect of the propagation through the beamline, which can be expressed in terms of a response function (see SI). Here, we assume that the propagation through the HHG gas target is linear, i.e. that nonlinear effects such as self-phase modulation and plasma generation do not strongly affect the pulse, as shown below. This fixes the waveform and CEP for the neon input pulse. For argon, we expect strong pulse reshaping in the HHG target, and therefore can not directly use the measured streaking waveform to obtain the input driving laser. Instead, we calculate the waveform from the neon input pulse with an additional amount of 260\,$\mu$m fused silica dispersion added to the spectral phase. This fixes the pulse envelope but not the CEP of the argon input pulse, since the relative CEP of the pulses in front of the wedges was not locked to the same value when switching gas species. The CEP of the argon pulse is fixed to the one whose focal field in the streaking chamber, after nonlinear propagating through the HHG gas target and the beamline, agrees best with the measured field, which, as we like to emphasize, also happens to be the one that optimizes the generation of isolated pulses in the simulations (see methods and SI for further details).

The experimentally measured XUV spectrum in argon is shown in Fig. \ref{Fig:Spectrogram} (h) and compared to simulations in (i). The experimental geometry of the spectrometer as well as the Zr-filter of 500\,$\mu$m thickness in front of the CCD camera has been taken into account in the simulations (see SI for details). The spectra have been normalized to the maximum for $\phi_0 +\pi/2$. A good agreement between the experiment and simulation is observed, only the amplitude at the maximum is slightly underestimated, but other details as the step-like increase of the spectrum for CEP=$\phi_0$ (red curves) at lower energies are reproduced. Furthermore, the XUV pulse intensity is shown in Fig. \ref{Fig:Spectrogram} f) and g) as black dashed lines. There is a slight overestimation of the trailing edge in the neon pulse shape which might be due to an overestimated weight of long trajectories in SFA compared to full quantum mechanical simulations\cite{Gaarde2008Review}, but overall there is an excellent agreement between simulations and experimental results.

\begin{figure*}[htb!]
	\centering
	\includegraphics[width=16.51cm]{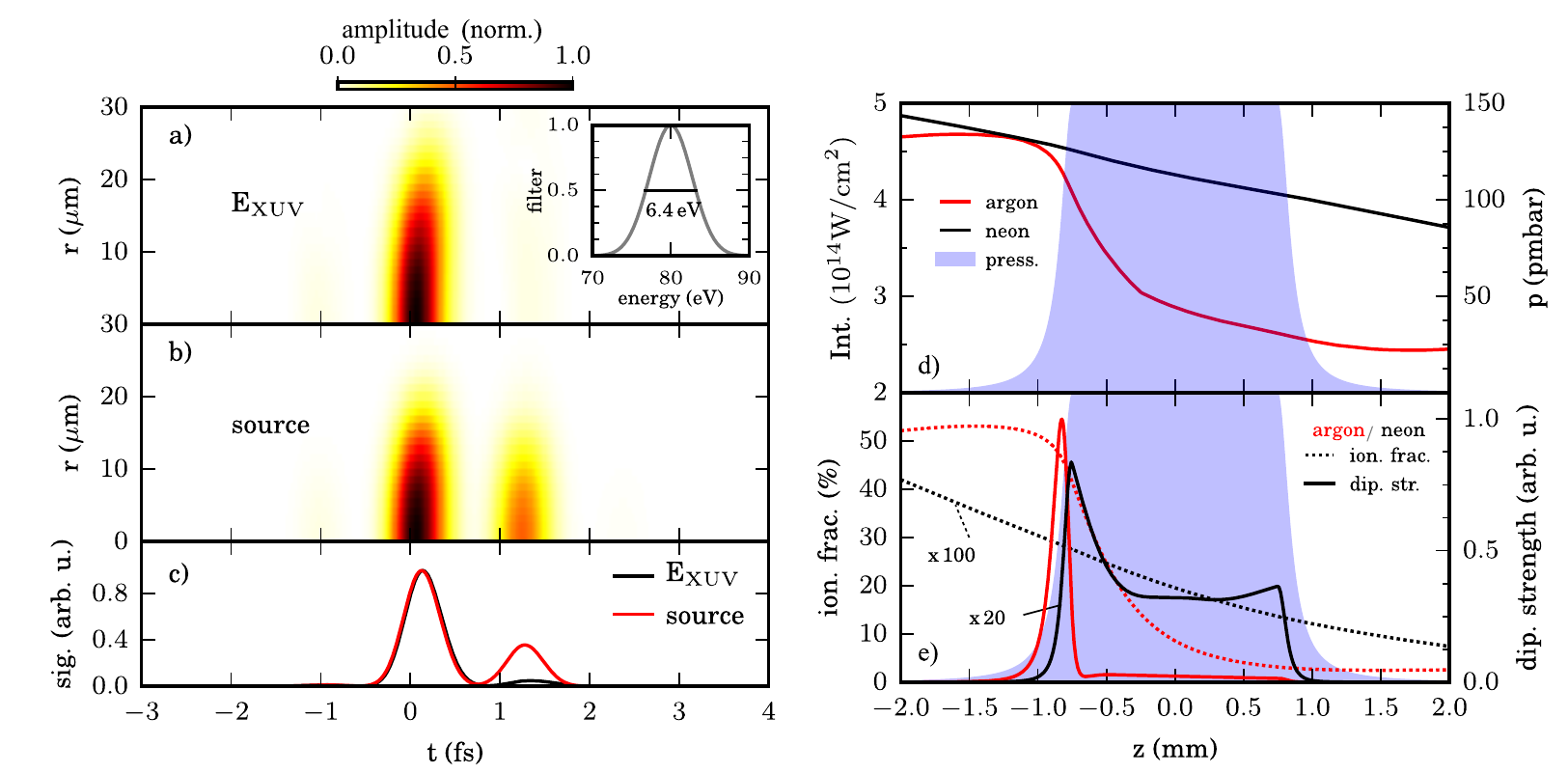}
	\caption{
		\label{Fig:Ionization_Intensity_Pol_XUV} Simulation of XUV generation in argon and evolution of the driving laser:  a) Spatiotemporal XUV pulse at the end of the target calculated within the energy window illustrated by the inset, which shows an isolated attosecond pulse. b) Spatiotemporal XUV source term within the above energy window integrated along the propagation axis assuming perfect phasematching. c) Comparison of the far-field XUV pulse at the position of the double mirror (solid black line) with the radially integrated source term from b) (red). The absence of the satellite pulse at around 1.3\,fs compared to the source term indicates that a transient phase-matching mechanism is present. d) Evolution of the peak intensity of the driving laser pulse on the propagation axis within the gas target for neon (black line) and argon (red line). The blue shaded areas here and in e) show the gas pressure distribution for both argon and neon. The propagation in argon is strongly influenced by the interaction with the gas. e) XUV source term (solid line) within the energy region specified in the inset of a) and the ionization fraction at the end of the pulse (dashed lines) for neon (black) and argon (red). The source term for argon is restricted to the entrance of the gas target. Note the different scaling for argon and neon.}
\end{figure*}

\subsection{XUV generation in the overdriven limit}

With the help of the simulations, we try to better understand the generation mechanism for isolated attosecond pulses in argon. Figure  \ref{Fig:Ionization_Intensity_Pol_XUV} shows the spatiotemporally resolved XUV electric field amplitude at the end of the HHG target within the reflection bandwidth of the XUV multilayer mirror. It is calculated from the HHG spectrum (including the phase) by multiplication with a Gaussian filter function with 6.4\,eV bandwidth centered at 80\,eV, as shown in the top panel, which closely resembles the multilayer mirror reflectivity. The resulting pulse, after Fourier transformation, shows a strong peak at around 0.1\,fs  and some very small, almost not discernible, satellite pulses at around -1.1\,fs and 1.3\,fs. The smooth radial profile seems to indicate that off-axis phase-matching does not play a major role here, as opposed to tighter focusing conditions\cite{Chen2017plasmadefocusingExp}. 

To see whether and how the XUV buildup is influenced by any transient phase-matching, we compare the XUV field amplitude to the time-resolved HHG source term in the same energy window. We calculate the latter for each spatial coordinate from the nonlinear polarization in the HHG target by the same energy filtering and Fourier transformation as above (see methods for further details). From this the XUV pulse that would approximately result for perfect time-independent phase-matching is determined by integrating the absolute value of the HHG source amplitude along the propagation axis. The result is shown in Fig. \ref{Fig:Ionization_Intensity_Pol_XUV} b) and together with a) gives a detailed information on the XUV generation process. In comparison to the XUV pulse, the HHG source term exhibits, besides the main pulse at around 0.1\,fs, another significant contribution about one half-cycle later, at around 1.3\,fs, which is strongly suppressed in the XUV pulse. The suppression of the second pulse must be a consequence of a time-dependent macroscopic phase-mismatch. As a side remark, we want to point out that the HHG source term is slightly more centered around the propagation axis compared to the XUV field, which is due to the neglect of diffraction. The suppression of the second pulse is further illustrated in c), which shows the XUV electric field (black solid line) and HHG source (red solid line) integrated along the radial axis.

\subsection{Phase-matching in the overdriven limit}

In order to understand how different aspects of the pulse propagation affect the XUV generation and buildup, we need to briefly revise the general expression for phase-matching. In the most general form, the phase-mismatch $\Delta k(\omega_q)$ of the $q$-th harmonic at frequency $\omega_q$ is given by the difference between the wavevector $k$ and the phase-gradient of the source-term\cite{Gaarde2008Review}: 
\begin{equation}
\label{Eq.Phasemismatch_Nat}
	\Delta k(\omega_q)=  k_{\omega_q}-\nabla \phi_{source}= k_{\omega} - q \cdot\nabla \phi_{IR}-\nabla \phi_{dipole}. 
\end{equation}
As shown above, the source-term can be further decomposed into a contribution from the driving laser field $\phi_{IR}$ and the intrinsic HHG dipole-phase $\phi_{dipole}$. The former contains the Gouy phase as well as the wavevector of the driving laser field, which depends on the gas pressure and plasma density. The latter term, which is absent in perturbative harmonic generation, depends both on the driving laser intensity $I$ and frequency $\omega_0$ and can be approximated by: $\phi_{dipole}=- \alpha_j U_{p}/\hbar \omega_{0}$\cite{Gaarde2008Review}, where $\alpha_j$ is an energy dependent proportionality constant which differs for long and short trajectories and $U_p[eV]=9.22\cdot \lambda[\mu m]^2 \cdot I[10^{14}W/cm^2]$ is the ponderomotive potential. Since the driving laser can lead to a significant increase (decrease) of plasma (neutral gas) density between different half-cycles which in turn might influence the pulse itself, all quantities additionally depend on time $\tau$. As we are interested in the XUV far-field close to the optical axis, and because there is additionally no indication for a particular off-axis contribution (see Fig. \ref{Fig:Ionization_Intensity_Pol_XUV} a), we restrict our analysis of the phase-mismatch in the gas target to the propagation axis ($r$=0) in longitudinal direction ($\vec{k}_{\omega_q} || \vec{e}_z$) and to the energy window around 80\,eV, unless stated otherwise.

For a driving laser pulse that is slowly varying, the gradient of the dipole contribution as well as the Gouy phase can be neglected, such that the phase-mismatch is given by the wavevector difference of the driving laser and HHG radiation, which is determined by the fraction of ionized gas atoms $\eta$: 
	\begin{equation}
	\label{Eq.phase-matching_plasma}
	\Delta k= q \cdot k_0\cdot \frac{\rho}{\rho_0}\cdot \Big[ (1-\eta)\cdot (n_{\omega_q}-n_{\omega_0})- \eta\cdot\frac{\rho_0 e^2}{2 \epsilon_0 m \omega_0^2}\cdot\Big(\frac{1}{q^2}-1\Big)\Big],
	\end{equation}

with the particle density $\rho$, the refractive index $n$ at the reference particle density $\rho_0$ and the electron mass $m$. For high harmonic orders $q$ the XUV contribution to the phase-mismatch can be neglected ($n_{\omega_q}\approx 1$ and $1/q^2\approx 0$), i.e. the generated XUV propagates with the speed of light $c$. We will always apply this approximation hereafter. With this assumption, independent of the pressure and harmonic order, the contribution of the neutral gas atoms is negative, while the plasma term is positive, and only a certain ionization fraction $\eta_{cr}$ leads to perfect phase-matching\cite{Popmintchev2009PhaseMatchingCutoffs}. In the tunneling regime, the plasma-density increases during each half-cycle of the driving laser pulse. The intensity at which this plasma density is reached at the main half-cycle of the pulse determines the highest XUV energy for which phase-matching can occur. This is the so-called phase-matching cutoff. If the intensity is further increased, phase-matching is achieved already before the main peak at lower XUV intensities. If the phase-mismatch in the preceding and following half-cycle is significantly different from zero, i.e. at high pressures, this transient phase-mismatch can be used for gating in isolated attosecond pulse generation\cite{Agostini1999optimizing, Balcou2003global,KapteynMurnane2006generation, Leone2009isolated, KapteynMurnane2009characterizing, KapteynMurnane2014generation}. For argon at a wavelength around 800\,nm the critical density is 3.8\% and is reached with our pulse at 3.3\,$\mathrm{10}^{\mathrm{14}}$W/cm$^2$, which would give a phase-matched cutoff of 70\,eV. However, in our experiment we observe an XUV cutoff of around 80\,eV. While plasma-induced phase-mismatch might play a role in the gating mechanism, our experiments are not consistent with a slowly varying driving laser pulse and we have to go beyond the simple picture of a non-varying driving laser pulse. 

This is further motivated by Fig. \ref{Fig:Ionization_Intensity_Pol_XUV} c) which shows the simulated maximum intensity of the pulse on the propagation axis for HHG argon (red) and neon (black). The pressure distribution is shown as blue shaded area. While in neon the intensity profile closely follows the Gaussian beam, consistent with our assumptions above, the pulses are strongly modified in argon. As soon as the pressure rises, there is a strong drop in intensity from 4.7$\cdot$ to 3$\cdot\mathrm{10}^{\mathrm{14}}$  $W/cm^2$ within the first few 100\,$\mu$m of the target followed by a slower decrease. This behavior can be explained by the different ionization of the two gases, which is illustrated in Fig. \ref{Fig:Ionization_Intensity_Pol_XUV} e). The fraction of ionized atoms (dashed lines) after the pulse has passed is more than a factor 100 higher at the beginning of the target for argon (up to 50\%) compared to neon (up to 0.4\%), which is due to the difference in ionization potentials ($I_{\mathrm{p},\mathrm{argon}}$=15.8\,eV and $I_{\mathrm{p},\mathrm{neon}}$=21.6\,eV) and the highly nonlinear ionization probability in the tunneling regime. At these pressures the plasma generation in argon influences the pulse intensity through absorption and defocussing, which leads to the strong intensity decay on-axis. Moreover, the pulse experiences a blueshift. We note that due to the geometry of our beamline, which blocks the driving laser on the propagation axis, this reshaping is only very weakly present in the measured streaking curve, as confirmed by a full simulation of the HHG process, beamline propagation and refocusing in the streaking focus (see SI for discussion). The high ionization probability also influences the XUV generation and leads to an increase in the HHG source term (solid line), which is proportional to the ionization probability amplitude. While the source term in the 80\,eV window in neon (black solid line) stays almost constant throughout the target, there is a strong peak at the target entrance for argon (red solid line) which decays within few 100\,$\mu$m as the intensity decreases and the HHG cutoff falls below the considered energy window. That means that for argon in the region where the XUV generation takes place, the pulses get strongly modified which in turn affects phase-matching. We therefore introduce here a straight-forward application of Eq.\ref{Eq.Phasemismatch_Nat} to describe the phase-mismatch:

\begin{equation}
\label{Eq.phasemismatch_num}
\Delta k(\tau,z) = -q \partial_z \phi_{IR}(\tau,z)\\+\alpha\cdot U_p/\hbar \omega_0 \cdot (\frac{\partial_z I(\tau,z)}{I} -3\cdot \frac{\partial_z \omega(\tau,z)}{\omega}),
\end{equation}
with the phase of the driving laser field $\phi_{IR}$, the instantaneous intensity $I(\tau,z)$, and radial frequency $\omega(\tau,z)$. All values are calculated in the co-moving reference frame of our simulation. This, together with the assumption that the XUV propagates with the speed of light, is the reason that the XUV wavevector contribution does not appear in Eq. \ref{Eq.phasemismatch_num}. Since we investigate the phase-mismatch close to the cutoff at 80\,eV we set $\alpha=3.2$. In principle for a fixed XUV energy, $\alpha$ is also dependent on the intensity and laser-frequency, which we, however, ignore for simplicity. As a consequence the phase-mismatch is split into two contributions, one from the driving laser pulse and the other from the XUV dipole. We like to point out, that by using the numerical calculation, the first term intrinsically contains atomic and plasma dispersion as well as any geometric phase. 

\begin{figure*}[htb!]
	\centering
	\includegraphics[width=16.5cm]{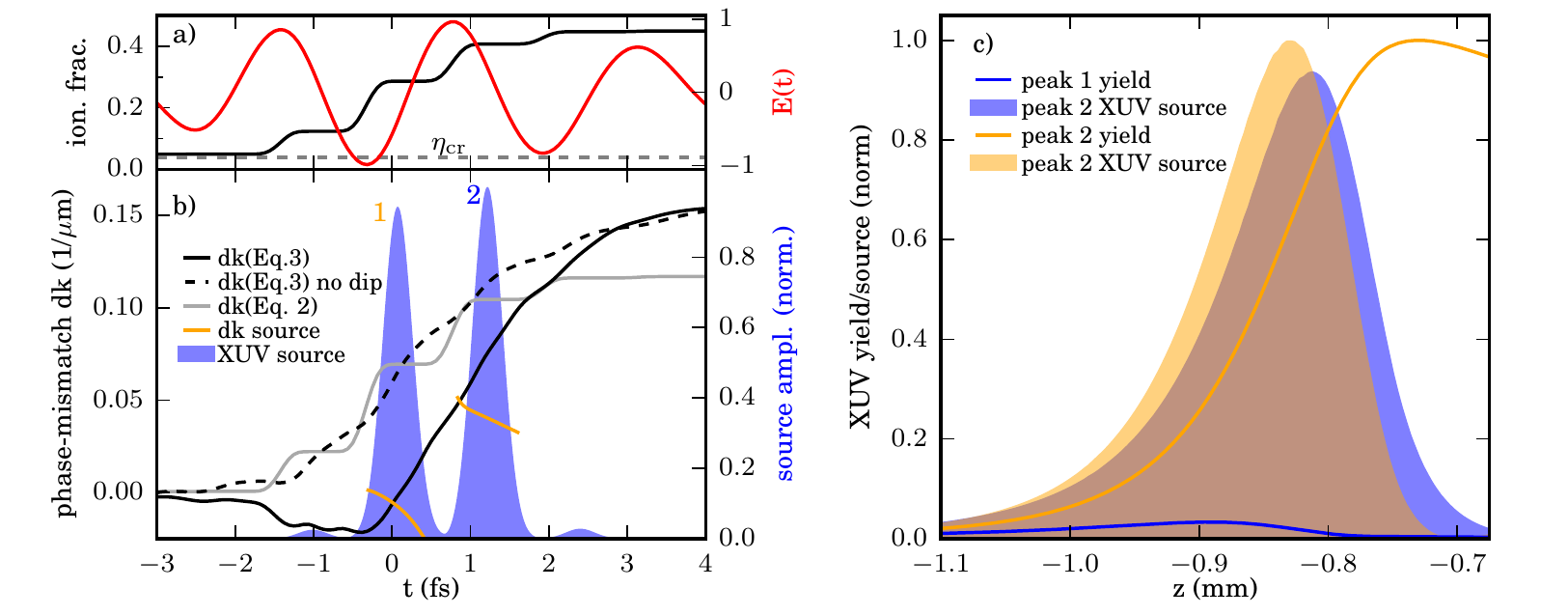}
	\caption{
		\label{Fig:phase_matching}On-axis transient phase matching and XUV generation in argon: a) The driving laser pulse (red) and the time-dependent fraction of ionized atoms (black). b) The on-axis phase-mismatch in the analytic expression (gray), the calculated phase-mismatch with the dipole contribution (black) and without (black dashed), and the numerical phase-mismatch calculated from the complex source term (orange) and the source term amplitude (blue shaded area) at z=-0.8\,mm. c) The on-axis source term amplitude (shaded area) and evolution of the XUV amplitude from the full simulation (solid line). The color indicates the first (orange) or second (blue) half-cycle centered around 0.1\,fs and 1.2\,fs, respectively, as shown in b).}
\end{figure*}

With this expression, we analyze the phase-matching for argon at the entrance of the HHG target (z=-0.8\,mm), at the maximum of the XUV source term. Figure \ref{Fig:phase_matching} a) shows the driving laser field (red solid line) and the fraction of ionized atoms (black solid line). Already at the beginning of the time window investigated here, the fraction of ionized atoms is 4.8\,\% and therefore above the critical ionization of 3.8\,\%. Figure \ref{Fig:phase_matching} b) shows the phase-mismatch calculated from different expressions (solid and dashed lines) and the XUV source term amplitude in the energy region around 80\,eV (blue shaded area). 
The XUV source term exhibits two main peaks of almost equal height from two different half-cycles at around -0.1\,fs and 1.2\,fs, similar to Fig. \ref{Fig:Ionization_Intensity_Pol_XUV} c). All phase-matching expressions show the general trend of an increase of the phase-mismatch as the plasma density increases over consecutive half-cycles, which qualitatively explains why the second peak is more strongly suppressed compared to the first one. However, the expressions neglecting the dipole contribution (solid gray and dashed black) exhibit a strong phase-mismatch also for the first peak and an overall small difference of the phase-mismatch of both peaks. In contrast, the expression from Eq. \ref{Eq.phasemismatch_num} (solid black line) that includes the dipole term shows a smaller phase-mismatch at the beginning of the pulse up to around 1.5\,fs and even a zero-crossing of the phase-mismatch for the first peak. This can be explained by closer examination of Eq. \ref{Eq.phasemismatch_num}. As the plasma density rises above $\eta_{\mathrm{cr}}$ the phase-mismatch due to the propagation of the driving laser increases (first term in Eq. \ref{Eq.phasemismatch_num}, see also Eq. \ref{Eq.phase-matching_plasma}). However since ionization, plasma formation and plasma defocussing generally lead on-axis to an intensity decay ($\partial_z  I<0$) and a blue-shift ($\partial_z \omega>0$), the last term is  negative and can counteract the plasma dispersion. Therefore phase-matching can be achieved even significantly above $\eta_{\mathrm{cr}}$ and also higher cutoff energies than predicted by the phase-matching cutoff are possible\cite{Chen2017plasmadefocusingExp}. Pulse reshaping therefore can enable phase-matching at least over small distances, as pointed out in Refs. \cite{Tempea2000NonadiabaticSelfPhasematching,geissler2000nonadiabaticPM,Yakovlev2007enhancedPM,Kaertner2011plasmadefocusing,Chen2017plasmadefocusingExp}, which is in contrast to the general notion for longer generation length that the driving laser pulse has ideally to remain unchanged (see e.g. \cite{KapteynMurnane2014generation}).

In order to evaluate this phase-matching expression, we compare them to the phase-mismatch numerically calculated from the XUV source term (solid orange line; see methods for details). Since we assume an XUV phase velocity of $c$ and are in a co-moving frame of reference, the numerical phase-mismatch is simply given by the derivative of the source term phase along the propagation axis. It is apparent that the numerically calculated value is only close to the full expression (Eq. \ref{Eq.phasemismatch_num}) and strongly deviates from the other two. However, the numerical expression shows the opposite tilt compared to the analytical expression. We believe that the main reason is that we assumed a constant $\alpha$, while in reality it increases from long to short trajectories, i.e. from earlier to later rescattering times. While this could be taken into account, it has to be considered that the HHG process is not instantaneous but takes around three quarters of a laser-cycle from ionization to rescattering. Therefore, the phase-mismatch can not be fully characterized by an expression that is calculated from instantaneous quantities of the driving laser and instead the whole evolution of the driving laser for the previous three quarters of an optical cycle would need to be taken into account. This could be achieved with expressions that contain integrals\cite{Tempea2000NonadiabaticSelfPhasematching, geissler2000nonadiabaticPM, Yakovlev2007enhancedPM}, however at the expense of drastically increased complexity. We conclude that Eq. \ref{Eq.phasemismatch_num} gives a decent estimate for the actual phase-mismatch.
The result of the phase-mismatch for the two different peaks of the XUV source term in Fig. \ref{Fig:phase_matching} b) for the XUV buildup on the propagation axis is shown in Fig. \ref{Fig:phase_matching} c). While the strength of the source term (shaded area) of both peaks is similar, the resulting XUV yield for the first peak (orange) is much higher than for the second (blue). This confirms yet again, that a transient gating mechanism is at play in our experiment, which is a consequence of the plasma-induced transient phase-mismatch discussed above. After the generation of the harmonic at the entrance, reabsorption in the target sets in. For the conditions in our simulations, XUV photons generated at the entrance of the target are transmitted with a probability of roughly 0.5 through the target for argon.

\subsection{Overdriven limit for different wavelengths and gases}

\begin{figure*}[htbp!]
	\centering\includegraphics[width=6.5in]{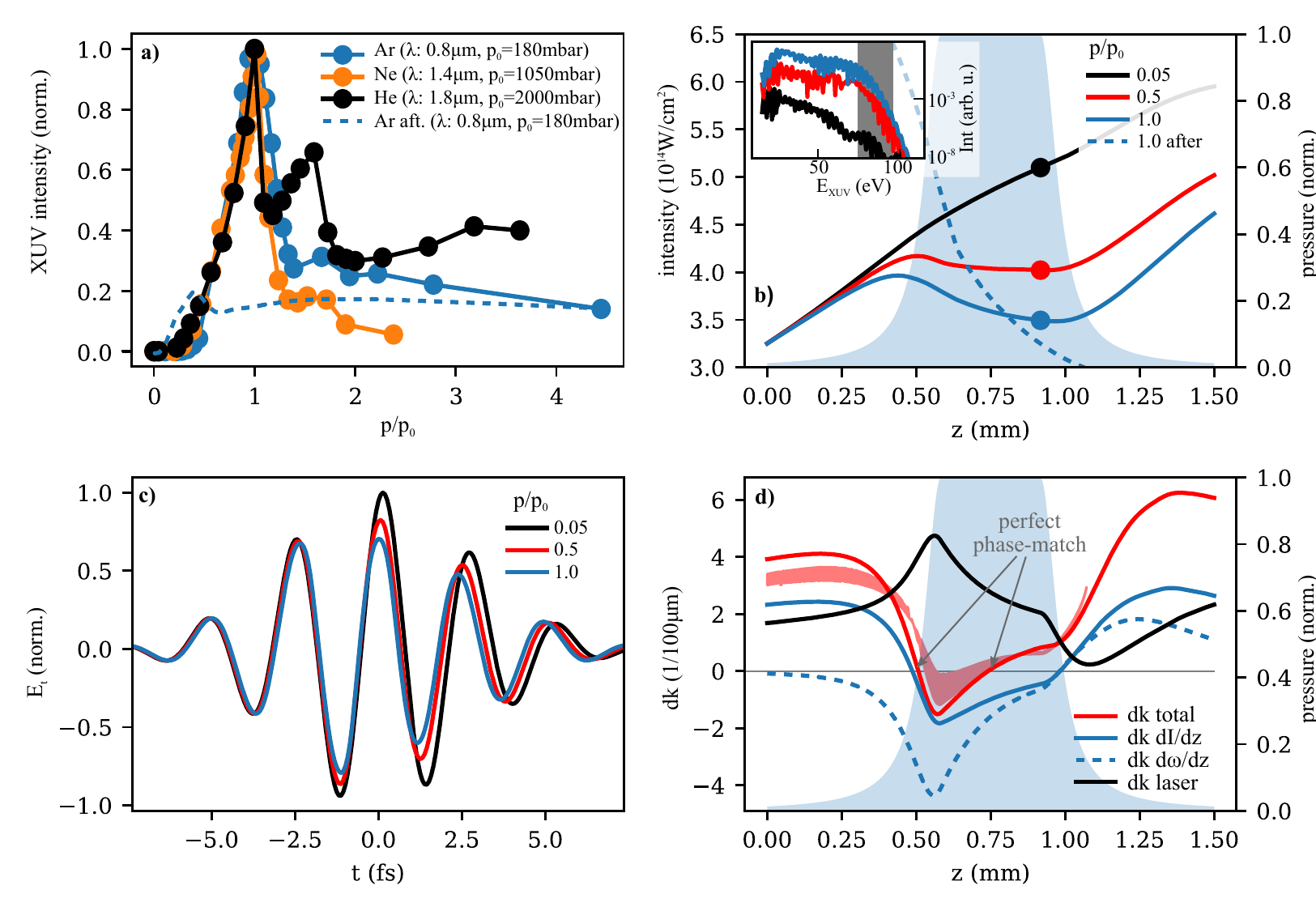}
	\caption{
		\label{Fig.pscanmain} High-harmonic generation dynamics in a tight focusing regime for different wavelengths and gases: a) the normalized far-field XUV flux in the cutoff region versus normalized pressure for a 1.75-cycle pulse at 800\,nm in argon (solid blue line, $p_0=180$\,mbar, 75\,eV-95\,eV, $w_0=20$\,$\mu$m), at 1400\,nm in neon (solid orange line, $p_0=1050$\,mbar, 250\,eV-320\,eV, $w_0=35$\,$\mu$m) and 1900\,nm in helium (solid black line, $p_0=2000$\,mbar, 560\,eV-700\,eV,$w_0=47.5$\,$\mu$m). Here the target is positioned in front of the focus. Additionally the flux in argon for a gas target positioned after the focus is shown as dashed blue line (upscaled by factor 10). b) on-axis maximum driving pulse intensity along the propagation direction for HHG in argon for different pressures p/$p_0$=0.05 (black), 0.5 (red) and 1.0 (blue) together with the curve for a gas target positioned after the focus. For the highest flux, the laser beam is strongly influenced by the interaction with the gas.The inset shows the XUV spectra together with the integration region (gray shaded area). c) Laser pulse at the end of the target as indicated by the dots in panel b). A clear intensity reduction and blue-shift of the trailing parts of the pulse are discernible. d) The different contributions to the total phase-mismatch (thick red line) on-axis for the most intense XUV burst for argon at $p=p_0$ calculated by Eq. \ref{Eq.phasemismatch_num}:  \textbf{dk laser} due to the propagation of the laser driving pulse (black line), \textbf{dk dI/dz} and \textbf{dk d$\omega$/dz} from the HHG dipole caused by the change of laser intensity (solid blue line) and the blue shift of the laser pulse (dashed blue line), respectively. Moreover, the numerically calculated total phase-mismatch within the half-cycle of the XUV burst is shown as red shaded area.}
\end{figure*}

So far, we have discussed how pulse reshaping affected the HHG phase-matching for our experiment and we have seen that the intensity decay and blue-shift can enable phase-matching. There are more general experimental conditions where these effects become important, namely HHG with longer wavelength driving lasers. These have gained interest in the recent years, fueled by the availability of high average power, few-cycle lasers in the infrared region and the promise of producing XUV pulses in the water window\cite{Popmintchev2012xrayreview, Biegert2016,Biegert2016soft_xray_continua,Kaertner2016,Tisch2018high}. There, the pulses are focused tightly into high gas pressure targets due to limited pulse energies and in this regime pulse reshaping due to ionization proved to be important in a recent study employing few-cycle pulses\cite{Tisch2018high}. We therefore performed another set of simulations under such conditions for different driving laser wavelengths and gases and studied the pressure dependence of the XUV yield near the conventional phase-matching cutoff. We used a 1.75-cycle (FWHM$_\mathrm{intensity}$) driving laser pulse with a CEP of 0 at a central wavelength of 0.8\,$\mu m$ for argon, 1.4\,$\mu m$ for neon and 1.9\,$\mu m$ for helium. The focal spot size of the Gaussian beam ($\omega_{0,Gauss}$) is taken to be 47.5\,$\mu m$ with a target length of 0.8\,mm and a pressure decay length of 0.15\,mm for a driving laser wavelengths of 1.9\,$\mu m$ and linearly scaled to for the other wavelengths. The target is placed in front of the focus such that the target entrance is at -0.5$\cdot$Rayleigh length. The reason why it is placed in front and not behind the focus as conventionally done, is discussed below. The intensity at the target entrance is chosen such that ten times the critical ionization would be reached at the central half-cycle of the pulse in free-space propagation.

The normalized XUV flux for different pressures and gas types/wavelengths is shown in Fig. \ref{Fig.pscanmain} a). The pressure axis is rescaled by the pressure of maximum flux ($p_{0,\mathrm{Ar}}=180 mbar$,$p_{0,\mathrm{Ne}}=1050 mbar$, $p_{0,He}=2000 mbar$). The XUV/x-ray flux was analyzed at a distance of 2\,m from the HHG target within a diameter of 5\,mm around the propagation axis and within an energy window of 75-95\,eV (argon), 250-320\,eV (neon) and 550-650\,eV (helium). The length scales are specific to our beamline but similar scales are relevant for other existing beamlines. Moreover, while slight variations occur with different parameters, we only aim at a qualitative comparison here. All three curves show a similar behavior. After a strong increase below the maximum, the flux drops again rapidly but then stabilizes at a level depending on the gas type of a few 10\% of the maximum flux. This is consistent with the calculation and analysis in Ref. \cite{Yakovlev2007enhancedPM}. There is an oscillation observable after the first maximum which becomes very prominent for helium. To explain it, we have to consider the phase-matching below. The approximately constant level at high pressures can be qualitatively explained by the increasing localization of the source contribution at the entrance of the target, as observed in the simulations above, and the transmission factors through the gas target for the respective energy window at $p_0$, which is around 0.5 for neon, 0.8 for argon and 0.98 for helium. For comparison, we show the flux for argon (dashed blue line upscaled by factor ten), when the target is placed at 0.5 Rayleigh length behind the focus at an increased intensity that corresponds to 15\,$\cdot\eta_{\mathrm{cr}}$. There, the maximum flux occurs at a lower pressure and is almost two orders of magnitude smaller.

Figure \ref{Fig.pscanmain} b) shows the maximum intensity along the propagation axis for argon at different pressures, $0.05\cdot p_0$ (black solid line), $0.5\cdot p_0$ (red solid line) and $1.0\cdot p_0$ (blue solid line) together with the normalized pressure distribution (blue shaded area). The inset shows the corresponding far-field XUV spectra and the analyzed energy region (gray shaded area). At low pressures, there is almost no influence of the gas on the pulse evolution and the intensity increases through the gas target. However, with higher pressures, the plasma generation leads to a limitation of the intensity increase and at even higher pressures to a decrease. At the pressure for maximum XUV flux, the intensity within the target is clamped to about 4.0$\cdot10^{14}W/cm^2$. After the target the intensity increases again, as the wavevector components away from the optical axis that have not been strongly affected by the plasma, are still focusing. This intensity clamping is similar to what has been observed in the simulations in Ref. \cite{Kaertner2011plasmadefocusing,Tisch2018high}. It also shows the importance of using realistic pressure profiles in the simulations. Comparing this to the intensity evolution when the target entrance is placed behind the focus at the maximum pressure $p_0$ (blue dashed line), one can see that the intensity drops very rapidly. This is due to the increased plasma production in front of the entrance of the gas target together with initially already diverging beam which strongly affects phase-matching. It therefore becomes clear why placing the target in front of the focus is favorable. The plasma defocussing and beam convergence counteract each other, which leads to an almost constant intensity within the gas target at high pressures. We want to point out that at the high pressures used in experiments( e.g. \cite{Biegert2016soft_xray_continua,Tisch2018high}), even at lower intensities corresponding to the critical ionization fraction $1\cdot\eta_{cr}$, we still observe significant intensity clamping and blue-shift in our simulations, and therefore believe that the non-adiabatic phase-matching mechanism analyzed here, might have played a key role. 

In order to see how the pulse is affected by the plasma dispersion, Fig. \ref{Fig.pscanmain} c) shows the pulse shape at the end of the gas target as indicated by the filled dots in Fig. \ref{Fig.pscanmain} b). While the leading part of the pulse is only slightly affected, the intensity clamping and the blue shift can be observed. While part of the intensity decay is also due to energy loss through ionization, the main effect is plasma defocussing, which affects longer wavelengths even more. On-axis this leads to an intensity decrease and blue-shift.

Finally, in Fig. \ref{Fig.pscanmain} d) the individual contributions to the on-axis phase-matching according to Eq. \ref{Eq.phasemismatch_num} for the peak of the main XUV pulse at $p_0$ are shown. The contribution of the driving pulse (black solid line) is positive throughout the target due to the strong-plasma contribution. The HHG dipole term due to the intensity change (blue solid line) reflects the derivative of the intensity. It is first positive and then changes sign slightly before the full pressure is reached and then increases again after the target. The dipole term due to the blue shift (blue dashed line) on the contrary is negative before and within the target and then switches to positive at the end of the target. Since the dipole terms counteract the driving laser contribution to the phase-mismatch, the overall phase-mismatch (red solid line) is close to zero within the gas target. Indeed there is a zero-crossing slightly before the entrance and then close to the middle of the gas target. We compare the semi-analytic to the numerical expression (red shaded area) and again very good agreement is observed. At the condition of maximum flux, shown here, while starting to be localized towards the target entrance, the source term still extends over the whole target length. In order to explain the oscillation after the maximum observed in Fig. \ref{Fig.pscanmain} a), we have to investigate what happens with rising pressure: the maximum of the XUV source term gets increasingly localized at the entrance and therefore the second zero-crossing of the phase-mismatch does not contribute to XUV buildup and at the same time, the XUV gets absorbed in the target. We note that while the phase-mismatch curve also slightly changes, it qualitatively stays the same in this pressure range. Therefore the flux drops rapidly after the maximum. As the pressure rises further, the source term peak eventually overlaps with the first zero-crossing of the phase-mismatch which leads to a small secondary maximum. Again, since we investigate the on-axis flux in the far-field, while for certain pressures, off-axis phase-matching might be favorable\cite{Vozzi2011PMhighenergyMIR,Chen2017plasmadefocusingExp}, we limit ourselves to the analysis given above.
Since under the pressure conditions of maximum flux and above, ionization and plasma-induced phase-matching changes drastically between different half-cycles for few-cycle lasers, transient phase-matching will play an important role, as discussed for our experiments above. Moreover, even though we restricted ourselves to the analysis of the longitudinal on-axis phase-matching, we believe that Eq. \ref{Eq.phasemismatch_num} can easily be generalized to 3D, if for example divergence angle resolved HHG spectra are studied.

\section{Conclusion}
\label{sec:conclusion}

We have presented an experimental and numerical study of the generation of isolated attosecond pulses in argon and neon through high-harmonic generation at 80\,eV. To our knowledge that is the highest photon energy for which isolated attosecond pulses in argon driven by laser pulses in the 800\,nm wavelength region have been reported and it is clearly above the classical phase-matching cutoff. Both, attosecond XUV and driving laser pulses were characterized through attosecond streaking. While the generation of IAPs in neon was consistent with amplitude gating, we find that ionization induced transient phase-matching is the underlying gating mechanism in argon. We observe in numerical simulations that the high ionization of the medium leads to a rapid decrease of the peak intensity, thereby limiting the HHG to the first few 100\,$\mu$m in argon. Analyzing the XUV source term, we find that due to the high ionization rate, the single atom HHG process is significantly enhanced compared to neon.  We compare different analytic phase-matching expressions and through comparison with the numerical calculation, we find that only the extended expression introduced here provides a quantitative description. We show that the rapid decrease of the intensity as well as the blue-shift significantly contribute to the phase-matching close to the XUV cutoff, allowing phase-matched XUV generation for a single half-cycle at significantly higher ionization fractions compared to slowly varying intensity conditions. 

We also perform simulations for other gas types and wavelengths for conditions which are relevant to other recent experiments and analyze the pressure-dependent XUV flux close to the cutoff. At the maximum flux phase-matching is strongly influenced through blue-shift and intensity decay due to plasma generation. We find that placing the target in front of the focus, allows significantly higher flux under these conditions, as the beam convergence and plasma defocussing balance each other. Moreover, our results show how pulse reshaping influences the phase-matching under high-pressure and strong focusing conditions. These findings have important implications for the analysis of current and future experiments that aim at XUV generation in the water-window and beyond, which in turn promises many interesting applications.

\section*{Methods}

\subsection{Experimental optimization of the HHG process}

 Experimentally, we start with HHG in neon by first optimizing the gas pressure for maximum XUV flux. Note that due to the feedback loop of the automated gas valve from the background pressure in the HHG chamber only a rough control is possible, however, our results are not dependent on the exact value of the pressure (see SI for further discussion). We then continued by fine adjusting the fused silica wedge insertion for maximum CEP-dependence in the cutoff region of the XUV spectrum. Subsequently, a streaking spectrogram in neon is recorded by inserting the double-mirror. We then switch to argon after evacuating the gas supply line. Besides gas pressure and dispersion, we also change the position of the focusing mirror as indicated in a) and try to optimize flux, stability and CEP-dependence in the XUV spectrum. Originally, as inferred from plasma generation in low-pressure air, the HHG gas target had been placed several mm behind the focus. With this procedure, we end up with a shift of the focusing mirror by about 5.5\,mm towards the target, an additional roughly 260\,$\mu$m of fused silica and similar flow rates. Afterwards, again the streaking spectrograms are recorded.

\subsection{Simulation model}

The simulation code is based on a model used in previous publications\cite{hogner2017generation}. The spot size on the mirror focusing into the HHG target was 7\,mm. The HHG target was assumed to have a thickness of 1.5\,mm filled with 150\,mbar of either argon or neon. The pressure gradient outside the target was taken from Ref. \cite{Tisch2018high}. From the focal length, pulse energy and waveform, as well as the the generated XUV spectra, we estimate a peak intensity of 5.35$\cdot10^{14}$W/cm$^2$ for the neon input pulse. This results in 4.8$\cdot10^{14}$W/cm$^2$ for argon. The shift of the focusing mirror (f=75\,cm) was taken care of by placing the entrance of the HHG gas target for neon 5.5\,mm after the target and at the focus for argon. The gas targets are modeled with a Lorentzian pressure decay along the propagation axis from the entrance and exit, as in previous publications\cite{Kaertner2011plasmadefocusing, Tisch2018high}.

The pulse obtained at the end of the HHG target is propagated through the beamline using the propagator of the Helmholtz equation in paraxial approximation. The beam is calculated at the position of the iris (d$_{\mathrm{HHG-iris}}$=1.3m), the Zr-filter that spatially separates XUV and driving laser (d$_\mathrm{iris-filter}$=0.3\,m), the focusing outer-mirror (d$_{\mathrm{HHG-double mirror}}$=2.0m, r$_{\mathrm{XUV mirror}}$=3mm, r$_{\mathrm{streaking mirror}}$=12.5mm, f=125mm) and finally in the focus (d$_{\mathrm{double mirror-focus}}$=133mm). The system is assumed to be optically centered in order to be able to employ cylindrical symmetry, which neglects that the focusing mirror is slightly tilted (please refer to the SI for more details).

\subsection{Calculation of the time dependent XUV source term}

In the paraxial approximation the propagation of a pulse is given by:

\begin{displaymath}
		\frac{\partial E}{\partial z}-i[k(\omega)-\omega/v_\mathrm{g}] E-\frac{i}{2 k(\omega)}\Delta_{\perp} E=\frac{i\omega}{2n(\omega)c}\frac{P}{\epsilon_0},
\end{displaymath}
where $k(\omega)$ is the wave-vector, $v_\mathrm{g}$ is the group-velocity, $\Delta_\perp$ is the Laplacian of the transverse coordinates, $n(\omega)$ is the refractive index, $\epsilon_0$ is the permittivity of free space and $P$ is the nonlinear polarization. Note, that we perform the calculation in the co-moving frame of reference. The term on the right hand side is the source term. In absence of dispersion (second term=0) and neglecting diffraction (third term=0), it directly describes the change of the electric field. For the time-dependent XUV-source term, we use the r.h.s. of the above equation, multiply it with $\mathrm{exp}\Big(-2\cdot \mathrm{ln}(2)\cdot [(\hbar \omega-80\mathrm{eV})/6.4 \mathrm{eV}]^2\Big)$ and perform an inverse Fourier transform. With source term intensity, the squared source term is denoted. When referring to the integrated source term, we mean the integration of the absolute value of the source term.

\bibliography{literature}

\section*{Acknowledgments}

We are grateful for support from the King-Saud University in the framework of the MPQ-KSU-LMU collaboration. We thank Ferenc Krausz for fruitful discussions. The MPQ/LMU group acknowledges support from the DFG via the Munich Centre for Advanced Photonics. JS acknowledges support from the Max Planck Society via the IMPRS-APS. MFK is grateful for support from Max Planck Society and from the DFG via the grant KL1439/10-1.

\section*{Author contributions statement}

M.F.K. and A.M.A.  conceived the experiment. J.S., B.F., W.S., I.L., H.A.M., A.M. K., and M.A. conducted the experiments. J.S. analyzed the data and developed the theoretical simulations based on code by M.H. and I.P.. C.J., N.G.K., and T.P-C. helped with the installation of the attosecond beamline and supported initial experiments. J.S. wrote the manuscript, which was reviewed by all authors.

\section*{Additional information}

The authors declare no competing financial interest.

%%%%%%%%%%%%%%%%%%%%%%%%%%%%%%%%%%%
%%%%%%%%%%%%%%%%%%%%%%%%%%%%%%%%%%%
%                         Supplementary Information                                       %
%%%%%%%%%%%%%%%%%%%%%%%%%%%%%%%%%%%
%%%%%%%%%%%%%%%%%%%%%%%%%%%%%%%%%%%
\pagebreak

\onecolumngrid
 \begin{center}
 	
	\textbf{\large Supplementary Information: Phase-matching mechanism for high-harmonic generation in the overdriven regime driven by few-cycle laser pulses}\\[.2cm]
	J. Sch\"otz$^{1,2,*}$, B. F\"org$^{1,2}$, W. Schweinberger$^{1,3}$, I. Liontos$^{3}$, H.A. Masood$^{3}$, A. M. Kamal$^{3}$, C. Jakubeit$^{2}$, N. G. Kling$^{1}$, T. Paasch-Colberg$^{1,3}$, M. H\"ogner$^{1,2}$, I. Pupeza$^{1,2}$, M. Alharbi$^{3}$, M.F. Kling$^{1,2,\dagger}$, A.M. Azzeer$^{3,\ddagger}$\\[.1cm]
		
	\email{matthias.kling@lmu.de}
	
	\email{azzeer@ksu.edu.sa}

	\email{johannes.schoetz@mpq.mpg.de}

	{\itshape ${}^1$ Physics Department, Ludwig-Maximilians-Universit\"at Munich, D-85748 Garching, Germany\\
		${}^2$Max Planck Institute of Quantum Optics, D-85748 Garching, Germany\\
		${}^3$Attosecond Science Laboratory, Physics and Astronomy Department, King Saud University, Riyadh, Saudi Arabia}\\[.4cm]
	${}^*$Electronic address: johannes.schoetz@mpq.mpg.de\\
	${}^\dagger$Electronic address: matthias.kling@lmu.de\\
	${}^\ddagger$Electronic address: azzeer@ksu.edu.sa\\
	{(Dated: \today)\\[1cm]}

	\end{center}
	
	\date{\today}
	
	\setcounter{equation}{0}
	\setcounter{figure}{0}
	\setcounter{table}{0}
	\setcounter{page}{1}
	\setcounter{section}{0}
	\renewcommand{\thepage}{S\arabic{page}} 
	\renewcommand{\thesection}{S\arabic{section}}  
	\renewcommand{\thetable}{S\arabic{table}}  
	\renewcommand{\thefigure}{S\arabic{figure}}
	\renewcommand{\refname}{References SI}
	\renewcommand{\th}{def}
	\flushbottom
	\maketitle
	
	\thispagestyle{empty}
	
	\section{Experimental Setup}
	%%%%% FIGURE LASER SETUP %%%%
	%%%%%%%%%%%%%%%%%%%%%%%
	\begin{figure*}[htbp!]
		\centering\includegraphics[width=15.7cm]{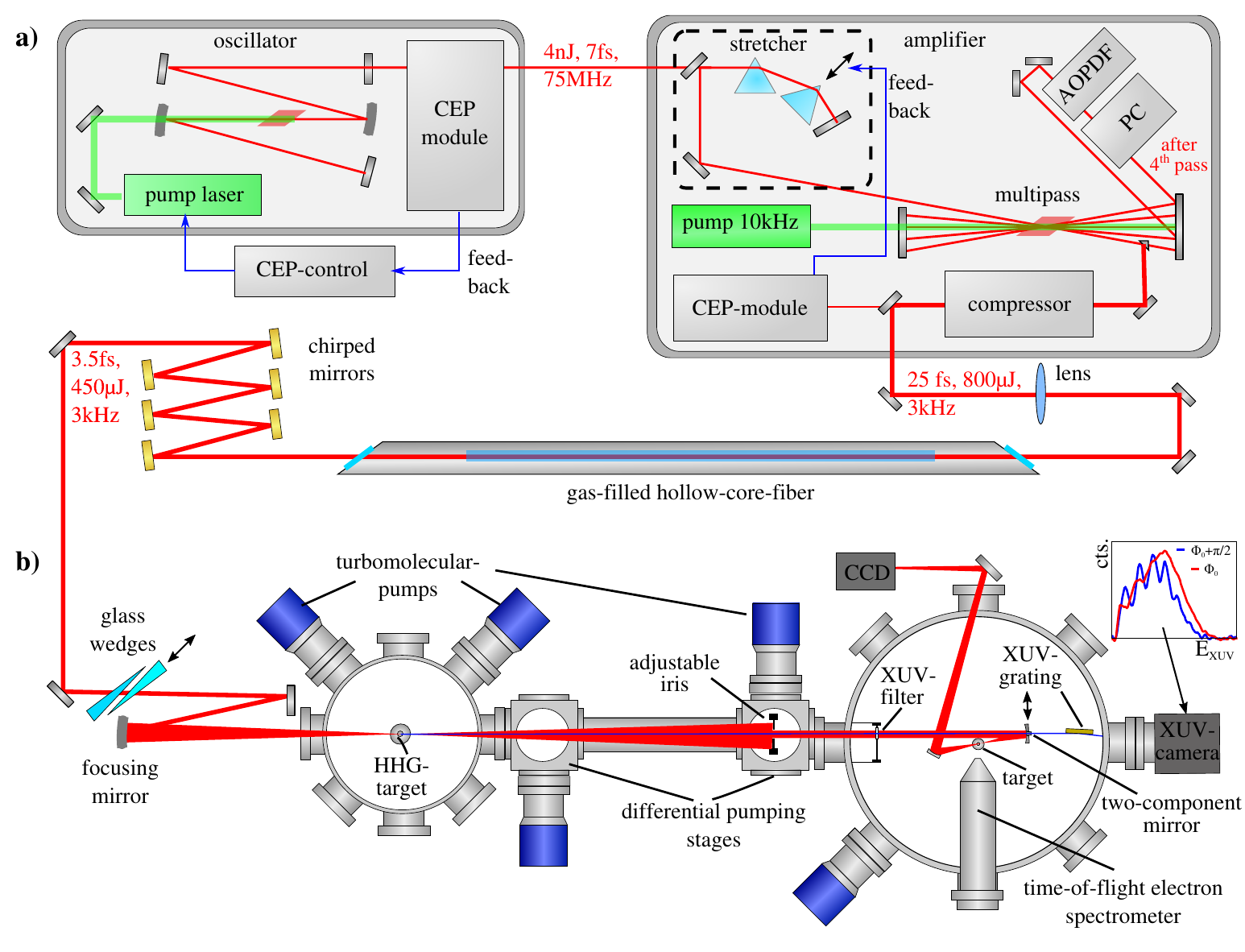}
		\caption{
			\label{Fig.Laser_Setup}Experimental setup: a) Laser system consisting of the oscillator, chirped-pulse amplifier, gas-filled hollow-core fiber and chirped mirror compressor. b) Vacuum beamline including HHG chamber, differential pumping chambers, and experimental chamber.}
	\end{figure*}
	
	The laser setup at the Attosecond Science Laboratory of the King Saud University is shown Fig. \ref{Fig.Laser_Setup} a). The pulses used in the experiment are delivered by a commercial laser system consisting of an oscillator (Spectra-Physics$^{\mathrm{TM}}$ ; rainbow$^{\mathrm{TM}}$ DFG seed) pumped by a solid-state diode laser (Coherent Inc.; Verdi V6 UNO). The carrier-envelope offset frequency is measured and controlled in a CEP module (Menlo Systems GmbH, XPS 800 Femtosecond Phase Stabilization) via a feedback to the pump power. The oscillator delivers 4\,nJ pulses at a repetition rate of 75\,MHz in the spectral range between 675\,nm and 930\,nm. The pulses are further amplified by a ten-pass chirped-pulse amplifier (Spectra-Physics$^{\mathrm{TM}}$ ; FEMTOPOWER$^{\mathrm{TM}}$  compact PRO HP/HR 3 kHz) which is pumped by pulsed solid-state diode lasers (Photonics Industries International Inc.; DM 30 Q-Switched DPSS Laser). After the fourth pass, a Pockels cell selects pulses at a repetition rate of 3\,kHz and an acousto-optic programmable dispersive filter (AOPDF; DAZZLER by FASTLITE) allows shaping of the pulse spectrum and phase. After the amplification, the pulses are compressed via a grating compressor to approximately 25\,fs with a pulse energy of 800\,$\mu$J. A small portion of the pulses is sent into a carrier-envelope phase (CEP) module (APS800 by Menlo Systems), which stabilizes long-term drifts via a feedback to one of the prisms in the stretcher. The pulses are spectrally broadened in a hollow-core fiber filled with neon to a width of 300\,nm centered at 750\,nm, supporting sub-two cycle pulses. The pulses are compressed by a set of chirped mirrors. Finally phase-stable few-cycle pulses with 350\,$\mu$J at 3\,kHz repetition rate are delivered to the experiment.
	
	The vacuum beamline is shown in Fig. \ref{Fig.Laser_Setup} b). The laser is focused with a concave mirror (f=75\,cm) into the HHG chamber. An iris is used to regulate the pulse power that reaches the experiment. XUV and fundamental beam are spatially separated by using a concentric Zr-foil filter and the focusing double mirror. The Zr-foil blocks the fundamental spectrum in the beam center that hits the inner mirror. The position of the outer and inner mirror can be changed by a closed-loop piezo stage (PI Hera 620). Other details are given in the main text. The gas flow through the HHG gas target is controlled via an automatic valve. The feedback to the controller is the background pressure within the HHG vacuum chamber. Unfortunately this does not permit to exactly determine the backing pressure within the HHG gas target. Is not possible to easily calibrate either since the dependence between HHG target backing pressure and the measured chamber pressure is nonlinear. Moreover, it depends on the entrance and exit hole sizes in the target, which changes over time. For our experimental parameters maximum flux in neon is usually achieved between 100-300\,mbar. We observed similar flow rates for argon and neon and therefore assumed a similar target pressure for both gas types of 150\,mbar. These pressures also give excellent agreement with experimental observations as shown in the main text. Moreover, as shown below in section \ref{sec:pressure_and_intensity}, the phase-matching mechanism that is strongly influenced by blueshift and ionization decay leading to a transient gating mechanism already becomes significant at much lower pressures. 
	
	\section{IR and attosecond pulse reconstruction}
	\label{sec::exp_res}
	
	\begin{figure*}[htb]
		\centering\includegraphics[width=6.25in]{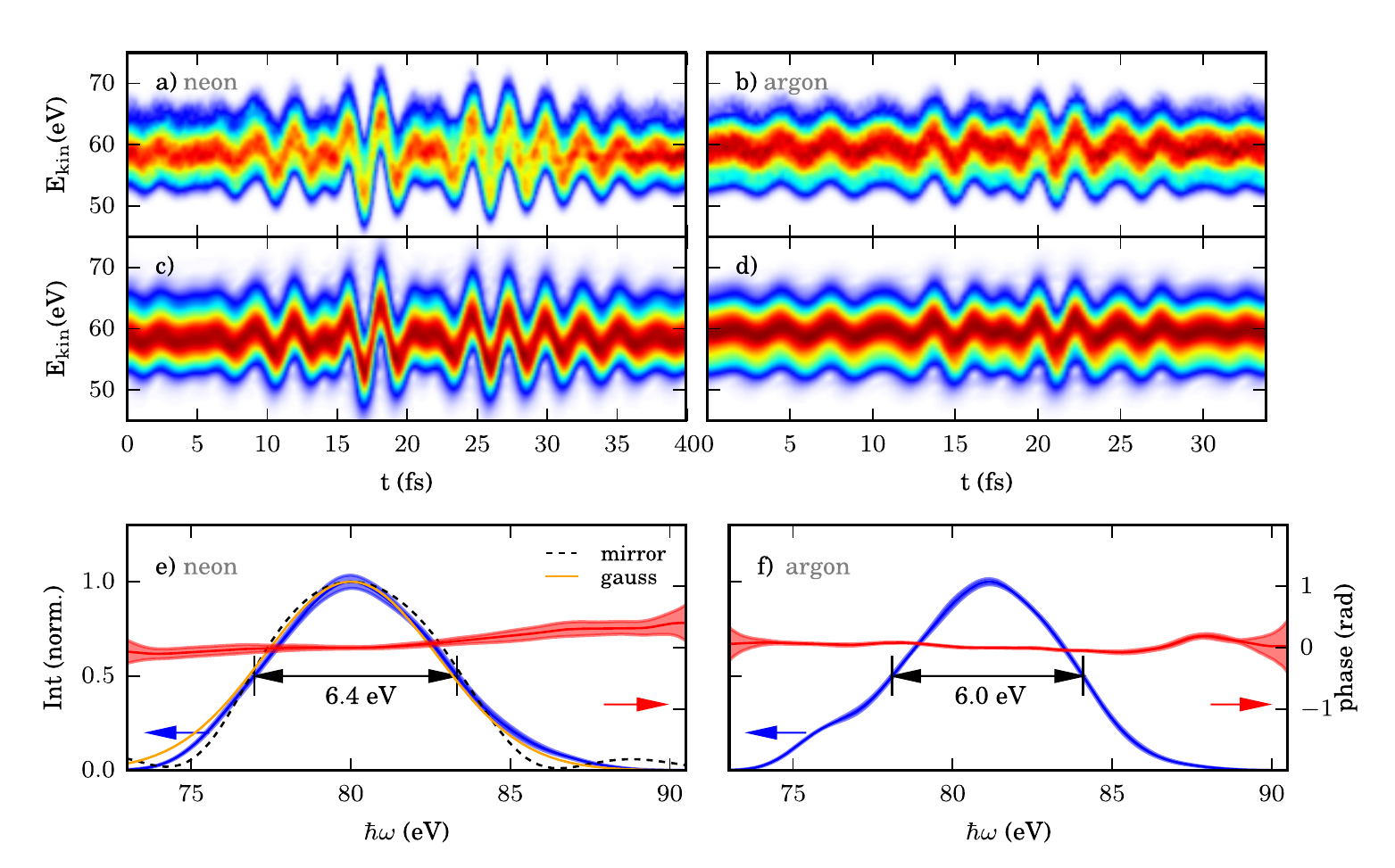}
		\caption{
			\label{Fig::XUV_reconstruction}
			Reconstruction of the XUV pulses using the ptychographic reconstruction algorithm: a) and b) filtered streaking spectrogram and c) and d) reconstructed spectrogram for neon and argon, respectively. The color scale is the same as in Fig. 2 a) and c) of the main text. In e) and f) the XUV spectral intensity, for neon and argon, respectively, is shown (blue line) together with the spectral phase (red line). The shaded area indicates the standard deviation of the last iteration in the reconstruction algorithm. In e) the normalized nominal reflectivity of the XUV multilayer mirror (maximum reflectivity approx. 42\,\%) and the Gaussian approximation (orange line) are also shown.}
	\end{figure*}
	
	In order to reconstruct the attosecond XUV-pulses we employed the ptychographic reconstruction algorithm described in Ref. \cite{Lucchini2015ptychographic}. It can simultaneously reconstruct the IR-pulse and has the advantage of being relatively easy to implement but relies on having a very good initial guess for the IR streaking field. The latter is given by the fit of the Gaussian functions to the electron spectrum for each delay step as described above. The initial guess for the XUV pulse is given by white noise in both spectral amplitude and phase. We found that the bare reconstruction algorithm introduced considerable high frequency components to the streaking field due to noise in the experimental measurements. We circumvented this by filtering out wavelength components below 100\,nm in the streaking field and, in contrast to the original algorithm, by averaging over the corrections over all delay steps in an iteration step instead of subsequently applying them (see Ref. \cite{Lucchini2015ptychographic} for details of the original algorithm). Furthermore, we dynamically adjusted the correction factor from 0.25 for the first 20 iterations to 0.1 for the remaining iterations to improve the quality and speed of convergence. With this approach, we achieved better convergence and considerably more consistent results than with a FROG-CRAB algorithm, judged on the basis of the spectrum of the reconstructed XUV and the frequency components of the driving field. It would be interesting to compare this algorithm in our experiment to more advanced reconstruction algorithms such as presented in Ref. \cite{Keathley2016GeneralizedATTOReconstruction}. This is, however, beyond the scope of this paper. The background counts in the spectrograms which lie in the quasi-homogenous, unstreaked spectral region below 55\,eV, attributable to both XUV radiation passing the Zr-filter and ionizing a gas atom on the way towards the XUV mirror and residual reflectivity of the XUV multilayer mirror at lower energies, had to be filtered out. This was achieved by applying a window function showing a quadratic increase (decrease) from 51.8\,eV to 54.3\,eV (65.8\,eV to 68.3\,eV) from 0 to 1 (1 to 0). The filtered spectrograms are shown in Fig. \ref{Fig::XUV_reconstruction} a) and b) for HHG generation in argon and neon, respectively, together with the reconstructed streaking spectrogram in c) and d) after around 200 iterations. In both cases the reconstruction is in excellent agreement with the input spectrograms. Since the algorithm reconstructs IR driving pulses and XUV pulses for each delay step individually, the final reconstructed pulses are calculated as the average of all delay steps. Please note that even though this provides a natural way to define the standard deviation, it is dependent on the correction factor. The reconstructed spectrograms shown here are calculated from the final averaged pulses. Fig. \ref{Fig::XUV_reconstruction} e) and f) show the spectrum of the XUV pulses (taking into account the ionization potential $I_\mathrm{p}=21.56$\,eV of neon). The XUV generated in neon shows an almost Gaussian intensity distribution with a full-width half-maximum of 6.4\,eV (orange line). The dashed line shows the normalized designed reflectivity of the mirror  (maximum reflectivity approx. 42\,\%). The other spectrum generated in argon shows a small shoulder at around 77\,eV and a reduced width of 6.0\,eV. This might indicate a slightly different generation mechanism, but could also be connected to the higher background in the streaking spectrogram for argon. The phase (red) in both cases is essentially flat. For simplicity, we assume in the main text a Gaussian reflectivity curve with 6.4\,eV FWHM. The reconstructed IR streaking pulses are shown as a black line in Fig. 2 d) and e) of the main text and practically coincide with the reconstruction from fitting the Gaussian functions which served as initial guess. 
	
	\section{Relation between HHG and streaking pulses}
	\begin{figure*}[htbp!]
		\centering\includegraphics[width=6.5in]{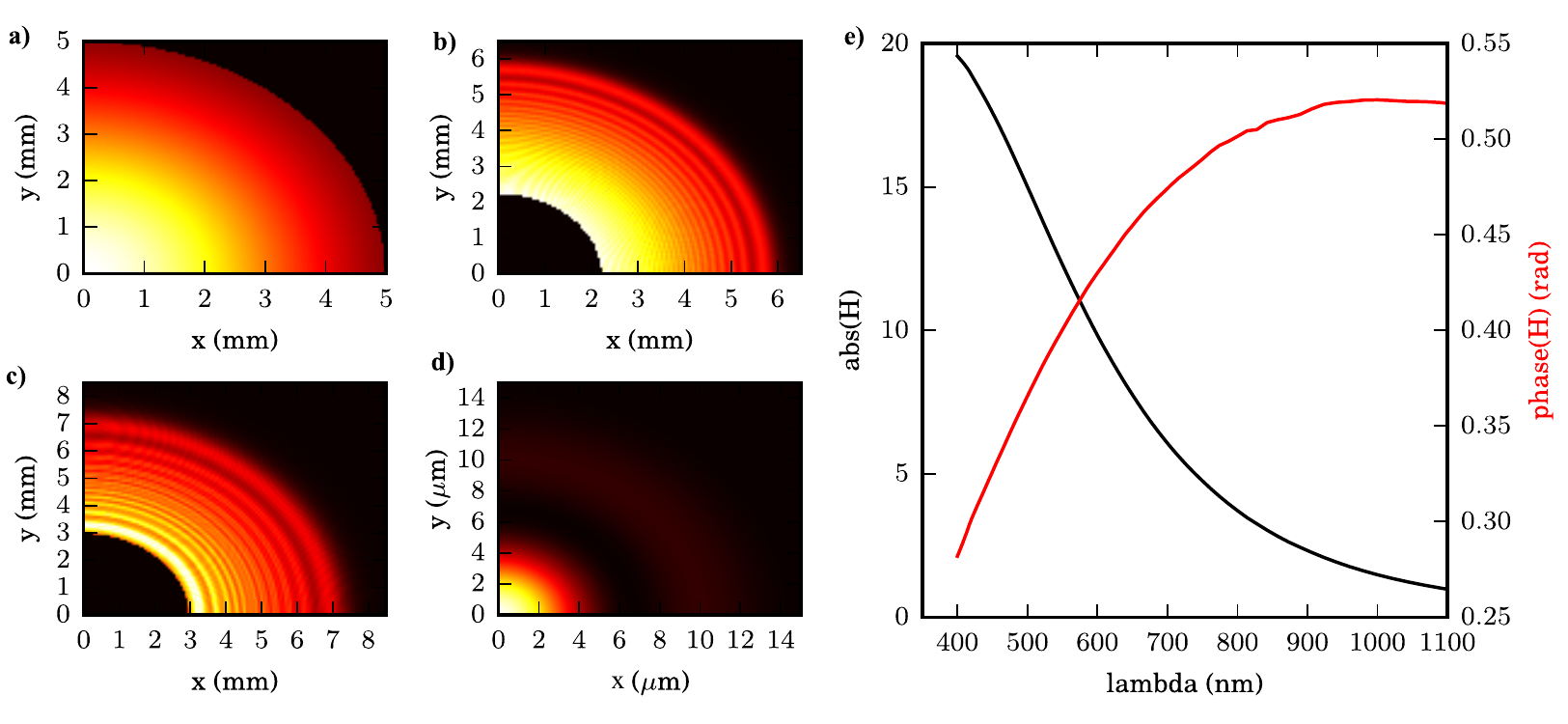}
		\caption{
			\label{Fig:modes_and_response_function}Calculation of the IR response: The modes in the beamline at the position of the iris (a), the Zr filter (b), the double mirror (c) and the focus (d). The images are calculated from the radial solution for the electric field. Any non-radially symmetric features stem from the interpolation onto the image grid. The absolute value (black) and phase (red) of the response function relating the electric field in the HHG target to the field in the streaking focus is shown in (e). Different iris openings and slight misalignments as well as the astigmatism found in the experiment affect more severely the absolute value than the phase.}
	\end{figure*}
	
	In order to assess the relation between the pulses in the HHG target and the ones in the streaking experiment, we numerically propagate the pulses through the beamline using the propagator of the cylindrically symmetric Helmholtz equation in the paraxial approximation:
	\begin{equation*}
		E(R,\Delta z)=\mathrm{exp}\Big(-j\frac{k_0\cdot R^2}{2 \Delta z}\Big)\int_{r_{\mathrm{min}}}^{r_{\mathrm{max}}} r  \mathrm{d}r \cdot E(r,0)\\ \cdot \mathrm{exp}\Big(-1j\frac{k_0\cdot r^2}{2 \Delta z}\Big) \cdot \mathrm{J}_0\Big(\frac{k_0 r R}{\Delta z}\Big),
	\end{equation*}
	where $E(R,\Delta z)$ is the electric field at radius R in the plane at $z=\Delta z$ that results from the electric field $E(r,0)$ at radii $r$ in the plane $z=0$, $k_0$ is the momentum vector and J$_0$ is the zeroth order Bessel function. In order to evaluate the above expression, we discretized it on linearly spaced radial grids for $r$ and $R$ and evaluated the integral through the trapezoidal rule. The approach is tested against the analytic solution of the Gaussian beam, and convergence is assured by doubling the number of grid points and comparing the results. Typically grids with 1500-2500 grid points are used. While the above method does not use the advantages of higher order integration schemes, it allows to efficiently incorporate irises and filters. The focusing mirror substrate (fused silica) is coated with 10\,nm of B$_4$C in order to achieve a reflectivity of about 10\,\% for the driving laser pulse. In the model, this is accounted for by using the reflectivity and phase for reflection from the thin-film at normal incidence (reflective index from Ref. \cite{larruquert2012selfB4C}).
	
	Panels \ref{Fig:modes_and_response_function} a)-d) show the modes for focal spot size of 50\,$\mu$m in the HHG target and a wavelength of 720\,nm at the position of the iris ((a), $z=$1.3\,m), the Zr-filter ((b), $z=$1.6\,m), the double mirror ((c), $z=$2.0\,m) and in the focus of the streaking experiment ((d),focal length$=$12.5\,cm) for an iris opening of 5\,mm. Clear diffraction rings are seen in Fig. \ref{Fig:modes_and_response_function} b)-d) which stem from clipping the beam at the iris and the filter.
	
	The on-axis response function which relates the pulses in the focus of the HHG chamber to the streaking experiment focus is depicted in e), again for an iris opening of 5\,mm. Since the short wavelength components are focused better, the absolute value of the response function (black line) strongly increases when decreasing the wavelength, which leads to an effective blueshift of the pulses measured in the streaking experiment. The relative phase (red line) increases from 0.30\,rad at 400\,nm to around 0.5\,rad at 800\,nm and stays almost constant up to 1100\,nm. This response is taken into account when determining the input pulse for HHG in neon from the measured pulses. Note, that the absolute value is bigger than unity for most of the spectrum, such that the resulting intensity in the streaking focus would be higher than in the HHG target. This is not observed in experiment, since the measured streaking intensities are on the order of $10^{12} W/cm^2$, which wouldn't allow HHG at 80\,eV photon energy in the HHG target. We found, that in order to resolve this discrepancy we have to take into account the astigmatism, i.e. the focusing under an angle at the double mirror. However to include the astigmatism, we have to give up cylindrical symmetry, which becomes memory expensive, and while the propagation via Fourier transform is still very simple, it suffers from the numerical reflection at the boundaries, which leads to strong oscillations in the phase. Therefore, we do neglect astigmatism so far. Sampling slightly outside of the focus either at $R>$0 or at different planes in propagation direction does mostly affect the amplitude of the response function. This means that while the phase, especially the CEP, is a good observable, the spectral amplitude has more uncertainty. The response function is taken into account when calculating the input pulse for HHG in neon from the measured streaking pulse, where only minor deviations from a free space propagation in the HHG target are expected.

	\section{Driving laser and pulse shape}
	
	\label{sec:driving_laser_pulses}
	\begin{figure*}[htbp!]
		\centering\includegraphics[width=6.5in]{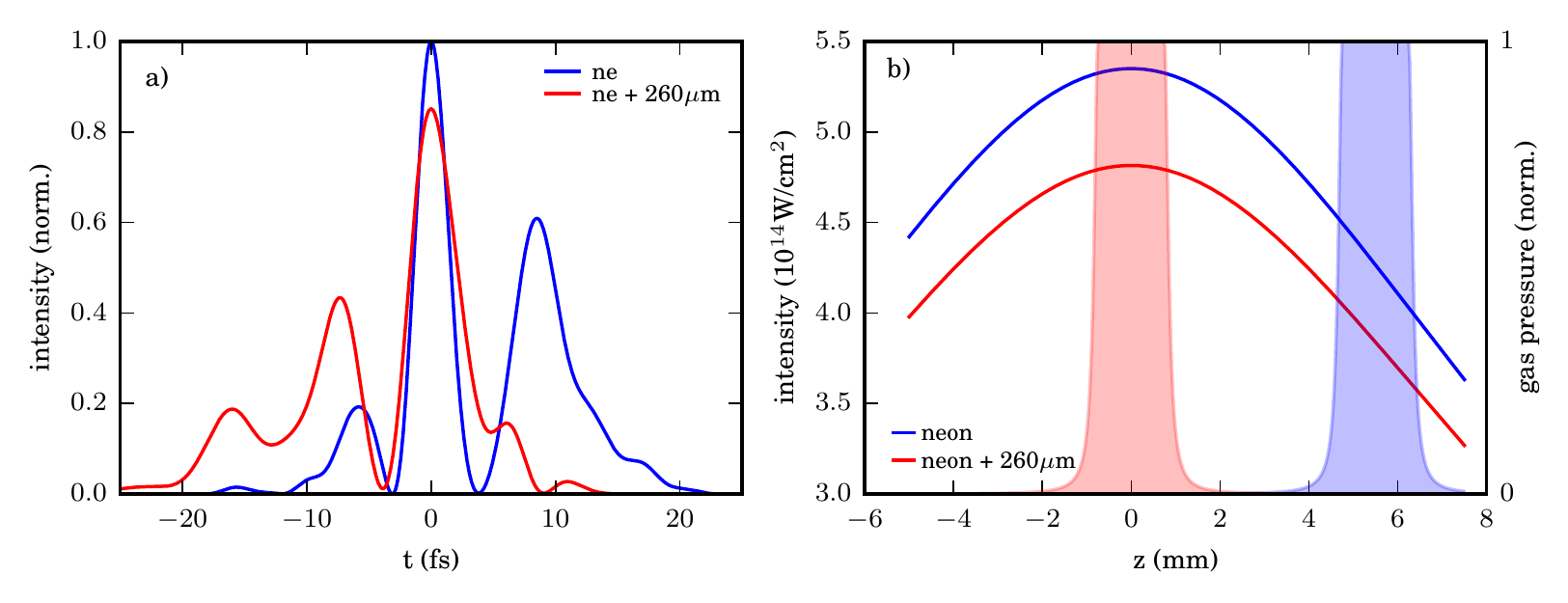}
		\caption{
			\label{Fig:neon_vs_argon_input_pulses} 
			Relation between the input pulse for HHG in neon and argon, which is modeled as the neon pulse with an additional 260\,$\mu$m of fused silica : a) The intensity of the neon input pulse (blue) and the input pulse for argon (red). b) The maximum intensity along the propagation axis for the two different input pulses (solid lines) and the assumed HHG target position indicated by the pressure profiles (blue shaded area for neon, red shaded area for argon).}
	\end{figure*}
	
	The input pulse for the XUV generation in argon is calculated from the neon input pulse by adding the dispersion of 260\,$\mu$m of fused silica. As can be seen in Fig. \ref{Fig:neon_vs_argon_input_pulses} a) the intensity of the pulse is decreased to slightly less than 90\,\%, such that the maximum intensity of the input pulse for argon (red line) is smaller than for neon (blue line). In order to still be able to reach intensities high enough for XUV generation at 80\,eV photon energy, the focus had to be shifted by roughly 5.5\,mm as was done in the experiment and an input peak intensity for neon of 5.35$\cdot$10$^{14}$W/cm$^2$ is assumed in the simulations (4.80$\cdot$10$^{12}$W/cm$^2$ for argon). This is shown in b) where the intensity along the propagation axis has been calculated for a Gaussian beam with 50\,$\mu$m focal spot size. The position of the gas target is illustrated by the gas pressure distribution (shaded areas).

	\section{Blueshift in argon and signatures of pulse reshaping in the streaking trace}
	
	\begin{figure*}[htbp!]
		\centering\includegraphics[width=6.5in]{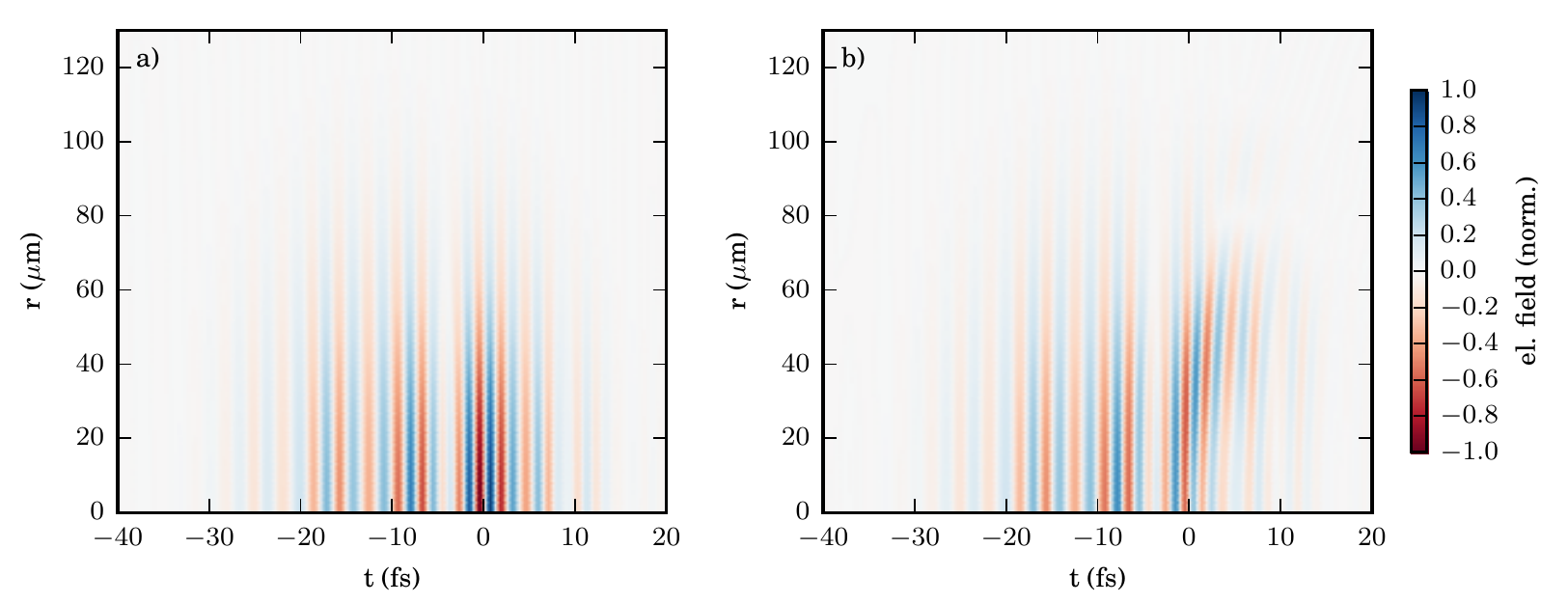}
		\caption{
			\label{Fig:driving_pulses_profiles} 
			Simulated driving pulse spatio-temporal profiles in argon at the start (a) and the end (b) of the HHG target. A strong modification of the driving pulses after the main peak at around 0\,fs is observed.}
	\end{figure*}
	
	In high-harmonic generation in argon the driving pulses undergo strong modifications as illustrated in Fig. \ref{Fig:driving_pulses_profiles} for a pressure of 150\,mbar. Panels a) and b) show the spatio-temporal pulse at the entrance and exit of the HHG target, respectively. In b) a strong decrease of the on-axis electric field amplitude after the main peak at 0\,fs is observed together with a blueshift. This is mainly due to ionization loss and plasma defocussing, with the latter being stronger for longer wavelengths. The modification of the pulse depends on the pressure in the target, as this controls the plasma density at the beginning of the target. 
	
	\begin{figure*}[htbp!]
		\centering\includegraphics[width=6.5in]{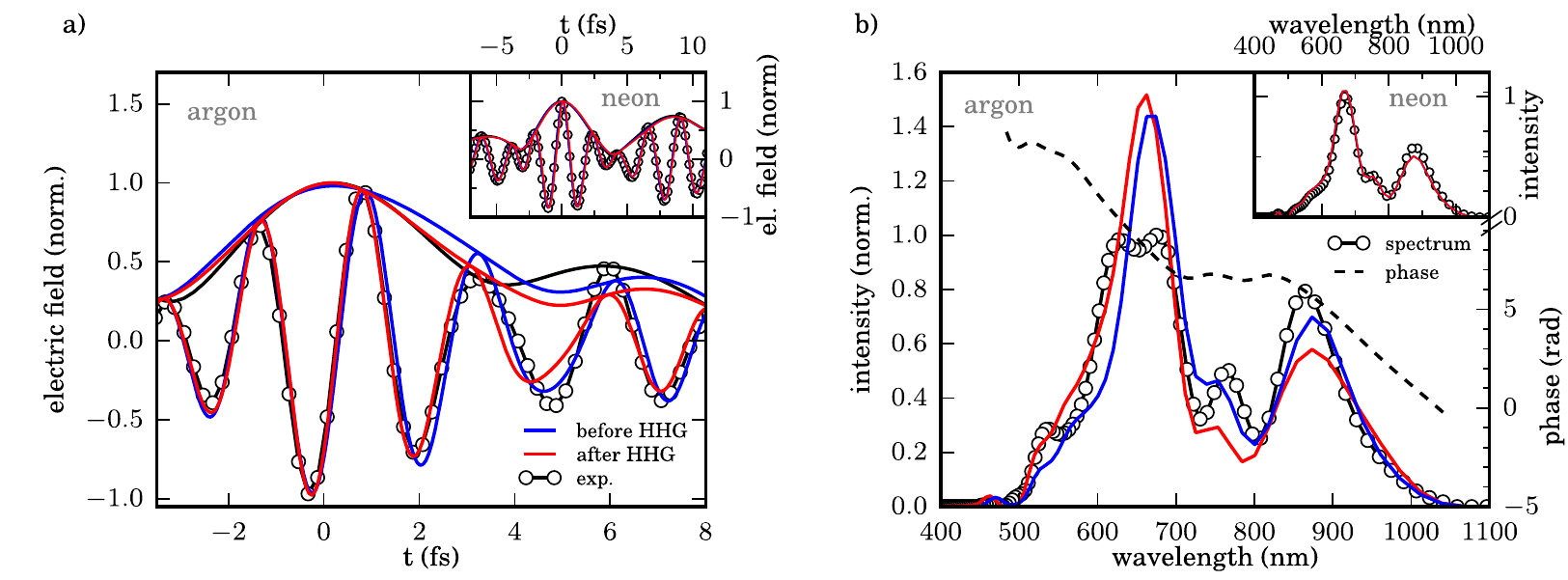}
		\caption{
			\label{Fig:streaking_fields_spectra} Signatures of pulse reshaping in the measured streaking field: Comparison of the laser waveform that was measured in experiment (white dots with black solid line), the simulated pulse in the streaking focus after propagation through the HHG target at 150\,mbar pressure and the beamline including focusing (red solid line), and the simulated pulse with no high-harmonic generation (blue solid line). Panel a) shows the comparison in the temporal domain, while b) covers the spectral domain. Main panels are for HHG in argon and the inset shows the neon results. A blueshift in a) in the measured waveform (black line woth white dots) is apparent in the comparison to the input pulse (blue) in the time window from 0 to +2\,fs. The simulation (red line), however, does not fully reproduce the experiments and is discussed in the text.}
	\end{figure*}
	
	Figure \ref{Fig:streaking_fields_spectra} a) shows the simulated laser electric field (red) in the focus of the streaking chamber in the temporal domain for HHG in argon (main panel) and neon (inset) at 150\,mbar. This is compared to the pulse that results from a target pressure of 0\,mbar (blue) and the experimentally measured waveform (solid black line and white dots). For neon all three waveforms agree almost perfectly. In the case of argon the overall waveform and especially the CEP agree quite well with the measured pulse, nevertheless a noticeable change can be seen upon propagation through the high-harmonic generation gas. Compared to neon the difference stems from the significantly higher third-order nonlinearity in argon\cite{wahlstrand2012high} as well as the different ionization rates, mostly determined by the ionization potentials (for neon $I_p$=21.6\,eV, for argon $I_p$=15.8\,eV), which are about a factor of 100 higher in the intensity range of 3 to 5$\cdot10^{14}{\mathrm{W}}/\mathrm{cm}^2$\cite{Tonglin2005JourPhysB} (see main text). Most notably, the rear part of the main peak from 0\,fs to 3\,fs is reduced, which leads to a shorter pulse duration, and a slightly increased oscillation frequency (blue-shift) is observed, which is consistent with the experimental observation, and is most likely a signature of the nonlinear pulse reshaping observed in the simulations. Between 3\,fs and 6.5\,fs, however, the simulated field that has undergone nonlinear interaction in the HHG target (red) seems to slightly overestimate the blue-shift and the intensity reduction compared to the experimental waveform. Nevertheless, we want to point out that the HHG process mainly takes place in the time window from -2\,fs to +2\,fs, where both experimental and simulated waveforms agree very well.
	
	The change of the waveform is also reflected in the spectral domain as can be seen in Fig. \ref{Fig:streaking_fields_spectra} b), which shows the spectrum of the streaking pulse for HHG in argon (main panel) and neon (inset). The spectra are normalized by the total intensity. Additionally, the spectral phase of the measured pulse is shown (right axis). For neon, again excellent agreement between the three curves can be seen. For argon, a clear bluehsift between the pulse propagated without gas (blue solid line) and after HHG (red solid line) can be seen. Indeed compared to the input pulse, the measured spectrum shows a similar blueshift and there is a decent agreement on the short wavelength side of the spectrum. However, while the simulated spectra show a peak around 650\,nm wavelength, the measured spectrum yields a saddle point. Additionally, there is an overestimation of the reduction of the red components, which might partly be due to the aforementioned point and the employed normalization method. 
	
	The overall trend of a blueshift and signatures of pulse reshaping are apparent in the experimental spectra. Some slight discrepancies between theory and experiment are present. There might be several reasons for this. First, the input mode is modeled as a Gaussian beam, which is a good approximation for the HCF output, but usually a slightly different mode with higher order diffraction rings is observed. This might have an influence, considering that in our geometry after the HHG target, the inner part of the driving laser pulse is blocked. Moreover, the change between gas species and optimization of the HHG process took about an hour. While the laser is usually very stable on that timescale, the supercontinuum generation in the HCF as a nonlinear process might be influenced by small drifts. Finally, the hole in the gas target is relatively small (100\,micron scale) and might block or diffract some part of the driving laser. Since we moved the target, this might lead to an additional change.
	
	\section{CEP scans of the HHG process in argon}
	
	\begin{figure*}[htbp!]
		\centering\includegraphics[width=6.5in]{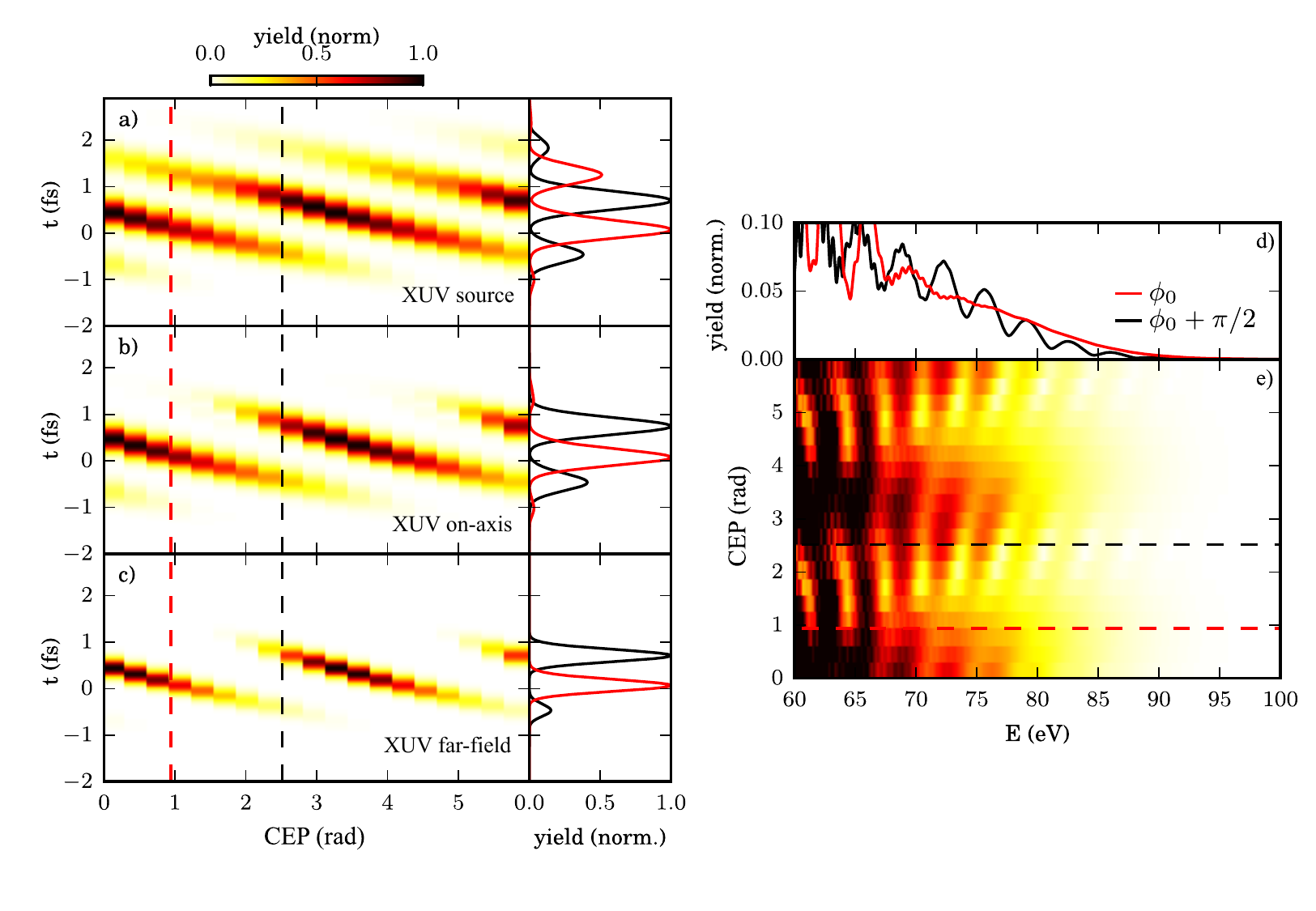}
		\caption{
			\label{Fig:CEP_scans}CEP-scan simulations: simulated time-dependence of the  source term radially and longitudinally integrated over the whole target (a), the XUV yield on axis at the end of the target (b) and the far-field XUV yield integrated over the inner mirror (c). The CEP $\phi_0$ used in the main text is shown as red dashed line and for comparison another CEP-value of $\phi_0+\pi/2$ (black) is shown. The panels to the right show the temporal profiles for the respective CEPs. The simulated far-field CEP-dependent XUV spectra are shown in d) together with the two lineouts at the two selected CEP values in e).}
	\end{figure*}
	
	In order to check the consistency of our simulations, we performed them also for different CEP values. Figure \ref{Fig:CEP_scans} a)-c) shows the temporal evolution of the XUV source term integrated over the radius and propagation axis (a), of the on-axis XUV at the end of the target (b), and the far-field XUV integrated over the inner mirror (c). The CEP-value $\phi_0$ that is used in the main text and that indeed optimizes the contrast of the isolated attosecond pulse generation, is indicated by the red dashed line and the respective lineout is drawn as a red solid line in the panels to the right. For comparison another CEP-value of $\phi_0+\pi/2$ is also shown with black dashed and black solid lines, respectively. As can be seen by comparing a) and b), while the XUV source term does not allow isolated pulse generation for any CEP, the contribution after the biggest peak (t$>$1\,fs) is strongly suppressed in the on-axis XUV yield due to phase-matching effects (as discussed in the main text). The far-field XUV yield resembles the on-axis contribution. For completeness, Figure \ref{Fig:CEP_scans} d) and e) show also the spectrally resolved CEP maps of the far-field XUV-yield and the individual spectra at the selected CEP values. Again, the positions of the selected CEPs are indicated by the red and black lines. 
	
	In Fig. 1 h) and i) of the main text, we compare the measured and simulated XUV spectra. In the simulation, the 500\,$\mu$m Zr-filter, which protects the CCD camera from the laser radiation, is considered, as well as the grazing incidence reflection from a plane round silver mirror at an incidence angle of around 88$^\circ$ and the height of the CCD chip. In the experiment, the mapping from the pixel position to energy was done using a quadratic fit to the position of the five clearly visible peaks, which are assumed to be spaced by 3.3\,eV. The offset energy is a free parameter, but agrees reasonably well with previous calibrations using an Al filter. Since the plane mirror, that is moved in and out of the beampath, is not on a closed-loop stage, the calibration is slightly different for each measurement. 
	
	\section{Pressure and intensity dependence of IAP generation in argon}
	\label{sec:pressure_and_intensity}
	\begin{figure*}[htbp!]
		\centering\includegraphics[width=6.55in]{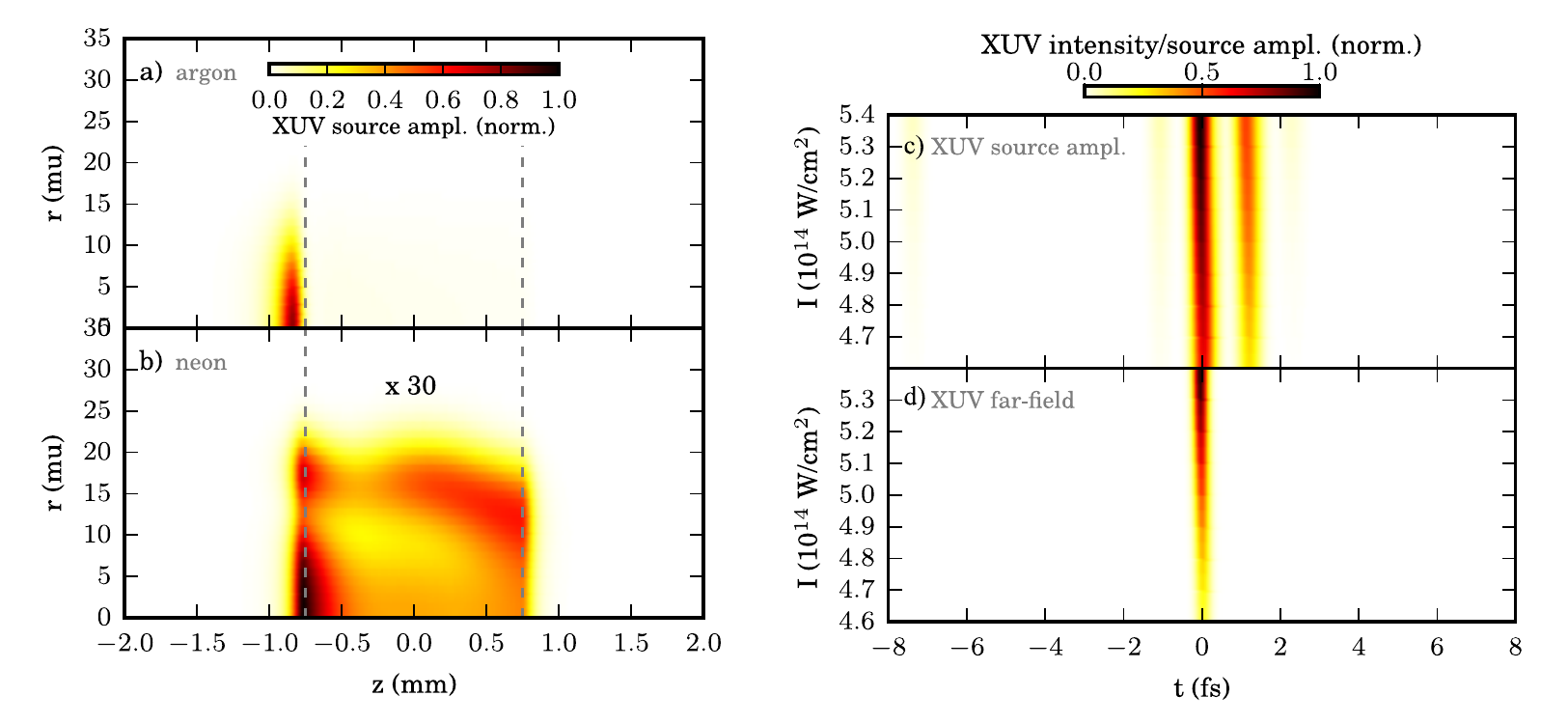}
		\caption{\label{Fig:spatially_resolved}
			Spatially resolved XUV source intensity in argon and neon (left) and the intensity dependence of the XUV source amplitude and far-field intensity in argon (right): Panels a) and b) show the spatially resolved XUV source intensity for argon and neon, respectively. The source intensity is integrated over the bandwidth of the mirror. Dashed lines indicate the limits between which the nominal pressure is reached. Panels c) and d) show the time-resolved on-axis XUV source amplitude integrated along the propagation axis and the far-field XUV intensity, respectively, for different intensities.}		
	\end{figure*}
	In this section, we compare the source term distribution in neon and argon and show that the mechanism observed in the main text does not critically depend on the intensity and pressure that we used for our simulations but plays a role for a broad parameter range. This is important since the pressure is not known very accurately in our experiments.
	
	Figure \ref{Fig:spatially_resolved} a) and b) shows the radially resolved XUV source intensity and integrated over the bandwidth of the mirror (Gaussian with 6.4\,eV intensity-FWHM at 80\,eV) for argon and neon, respectively. The driving pulse intensity and target position is the same as in the main text and as described in the section above. In both cases the source intensity radially extends to about 15-20\,$\mu$m.  However, in propagation direction, while in the case of neon the source extends over the whole target, in argon it is limited to the first few 100\,$\mu$m. The substructure in the XUV source intensity in neon stems from the fact that the XUV dipole strength does not monotonically decay with intensity, but may rise again once the XUV cutoff hits the considered energy window\cite{Corkum93threestep}.  
	
	Figure \ref{Fig:spatially_resolved} c) and d) shows the time-resolved on-axis XUV source amplitude integrated along the propagation axis and the far-field XUV intensity, respectively, for different driving laser intensities. The same spectral filtering and Fourier transformation as described in the main text has been performed. The overall yield increases with intensity. While the source amplitude shows a strong contribution at around 1.1\,fs, it is absent from the far-field intensity due to phase-matching for all the considered intensities. In conclusion, the phase-matching gating works over a broad intensity window.
	
	\begin{figure*}[htbp!]
		\centering\includegraphics[width=6.55in]{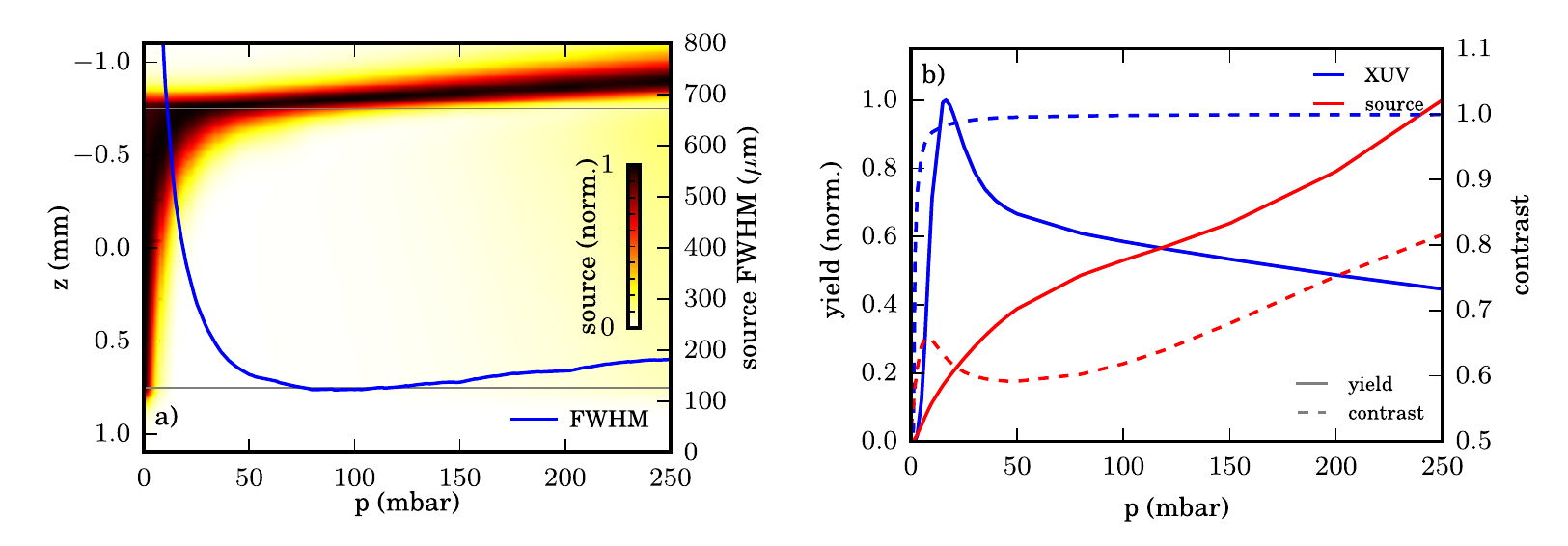}
		\caption{\label{Fig:pressurescansource}
			Spatially resolved XUV source intensity in argon and neon (left) and the intensity dependence of the XUV source amplitude and far-field intensity in argon (right): Panels a) and b) show the spatially resolved XUV source intensity for argon and neon, respectively. The source intensity is integrated over the bandwidth of the mirror. Dashed lines indicate the limits between which the nominal pressure is reached.}
	\end{figure*}
	We performed simulations for various pressures in order to evaluate the robustness of the phase-matching mechanism. Figure \ref{Fig:pressurescansource} a) shows the z-dependence of the radially integrated normalized source intensity. While at very low pressures, the source term extends over the whole target (gray lines indicate the region between which the nominal pressure is reached), it gets rapidly localized at the entrance of the target for higher pressures. Above approximately 80\,mbar, the source term reaches its maximum already slightly in front of the target, i.e. before the pressure has reached the nominal value. The full-width at half-maximum (FWHM) of the source term intensity along the propagation axis is also shown (blue line). Already above roughly 15\,mbar the FHWM drops below 500\,$\mu$m, meaning that the generation is already localized to within a fraction of the target length. For pressure higher than 50\,mbar the FHWM lies below 200\,$\mu$m. Once the maximum moves out of the target, the FHWM slightly increases again.
	
	As can be seen in Fig. \ref{Fig:pressurescansource} (b), the total XUV yield on the inner mirror (blue solid line) at first grows dramatically with pressure and reaches a sharp maximum at around 20\,mbar. After a drop up to about 50\,mbar to roughly 70\,\% of the maximum yield, it decreases only slowly with increasing pressure. The contrast between the main XUV pulse and other pre- and postpulses is shown as blue dashed line. Starting from a value of 0.5 at low pressures, indicating the presence of a double pulse, it rapidly reaches a value of 0.9 above 4.5\,mbar, 0.98 at the maximum of the yield at 20\,mbar and then slowly approaching 1.0 for higher pressures. As discussed above, the reason is that the phase-mismatch of the first peak is much lower than that of any of the following peaks, as it is growing with the fraction of ionized atoms. This is further illustrated by comparing the contrast of the far-field XUV yield (blue dashed line) to the contrast of the source term (integrated over $z$ and $r$). At very low pressures, both contrasts are at around 0.5 since the plasma density does almost not influence the propagation. Once the pressure is increased, the contrast of the source term rises slightly to around 0.6 from pressures of a few mbar and stays there until up to 150\,mbar, above which it slowly increases to 0.8 at 250\,mbar. The consecutive peaks strongly present in the source term are thus suppressed.
	
	%\bibliography{literature}

\end{document}